# Structural pathways for ultrafast melting of optically excited thin polycrystalline Palladium films

Jerzy Antonowicz[1]*, Adam Olczak[1], Klaus Sokolowski-Tinten[2,3], Peter Zalden[4], Igor Milov[5,6,7], Przemysław Dzięgielewski[1], Christian Bressler[4], Henry N. Chapman[6,8,9], Michał Chojnacki[10,11], Piotr Dłużewski[10], Angel Rodriguez-Fernandez[4], Krzysztof Fronc[10,11], Wojciech Gawełda[12,13,14], Konstantinos Georgarakis[15], Alan L. Greer[16], Iwanna Jacyna[10], Robbert W.E. van de Kruijs[5], Radosław Kamiński[17], Dmitry Khakhulin[4], Dorota Klinger[10], Katarzyna M. Kosyl[10], Katharina Kubicek[4], Kirill P. Migdal, Roman Minikayev[10], Nikolaos T. Panagiotopoulos[16], Marcin Sikora[18], Peihao Sun[19], Hazem Yousef[4], Wiktoria Zajkowska[10], Vasily V. Zhakhovski, and Ryszard Sobierajski[10]*

[1]Faculty of Physics, Warsaw University of Technology, Koszykowa 75, 00-662 Warsaw, Poland.
[2]Faculty of Physics, University of Duisburg-Essen, Lotharstrasse 1, 47048 Duisburg, Germany.
[3]Center for Nanointegration Duisburg-Essen (CENIDE), University of Duisburg-Essen, Lotharstrasse 1, 47048 Duisburg, Germany.
[4]European XFEL, Holzkoppel 4, 22869 Schenefeld, Germany.
[5]Advanced Research Center for Nanolithography (ARCNL), Science Park 106, 1098 XG Amsterdam, The Netherlands.
[6]Deutsches Elektronen Synchrotron DESY, Center for Free Electron Laser Science CFEL, Notkestr. 85, 22607 Hamburg, Germany.
[7]Industrial Focus Group XUV Optics, MESA+ Institute for Nanotechnology, University of Twente, Drienerlolaan 5, 7522 NB Enschede, The Netherlands.
[8]Department of Physics, Universität Hamburg, Luruper Chaussee 149, 22761 Hamburg, Germany.
[9]Hamburg Centre for Ultrafast Imaging, Luruper Chaussee 149, 22761 Hamburg, Germany.
[10]Institute of Physics Polish Academy of Sciences, al. Lotników 32/46, 02-668 Warsaw, Poland.
[11]International Research Centre MagTop, Institute of Physics, Polish Academy of Sciences, Al. Lotników 32/46, 02-668 Warsaw, Poland.
[12]Department of Chemistry, Universidad Autonoma de Madrid, Ciudad Universitaria de Cantoblanco 28049 Madrid, Spain.
[13]IMDEA Nanociencia, Calle Faraday 9, 28049 Madrid, Spain.
[14]Faculty of Physics, Adam Mickiewicz University, ul. Uniwersytetu Pzonanskiego 2, 61-614 Poznan, Poland.
[15]School of Aerospace, Transport and Manufacturing, Cranfield University, Cranfield, MK43 0AL, UK.
[16] Department of Materials Science & Metallurgy, University of Cambridge, Cambridge, CB3 0FS, UK.
[17]Department of Chemistry, University of Warsaw, Żwirki i Wigury 101, 02-089 Warsaw, Poland.
[18]Academic Centre for Materials and Nanotechnology, AGH University of Krakow, Al. A. Mickiewicza 30, 30-059 Krakow, Poland.
[19]Dipartimento di Fisica e Astronomia "Galileo Galilei", Università degli Studi di Padova, Padova 35131, Italy.

*Corresponding author. Email: jerzy.antonowicz@pw.edu.pl, ryszard.sobierajski@ifpan.edu.pl

**Abstract**
Due to its extremely short timescale, the non-equilibrium melting of metals is exceptionally difficult to probe experimentally. The knowledge of melting mechanisms is thus based mainly on the results of theoretical predictions. This work reports on the investigation of ultrafast melting of thin polycrystalline Pd films studied by optical laser pump – X-ray free-electron laser probe experiments and molecular-dynamics simulations. By acquiring X-ray diffraction snapshots with sub-picosecond resolution, we capture the sample's atomic structure during its transition from the crystalline to the liquid state. Bridging the timescales of experiments and simulations allows us to formulate a realistic microscopic picture of melting. We demonstrate that the existing models of strongly non-equilibrium melting, developed for systems with relatively weak electron-phonon coupling, remain valid even for ultrafast heating rates achieved in femtosecond laser-excited Pd. Furthermore, we highlight the role of pre-existing and transiently generated crystal defects in the transition to the liquid state.

# Introduction

Melting (1) is one of the most commonly observed but still not fully understood phase transitions. The problem of how crystalline solids melt and what determines the temperature at which the transition happens has fascinated scientists for over a century, and understanding of specific mechanisms underlying melting is still evolving. Of numerous theoretical criteria for melting, the most prominent are those proposed by Lindemann (2) and Born (3). The Lindeman criterion relates melting to a vibrational instability, stating that melting occurs when the root-mean-square vibrational displacement of atoms in a crystal exceeds a critical fraction of the inter-atomic spacing (4). The value of the critical fraction is not constant and can take values in the range of 0.05–0.2, depending on the type of interatomic interaction and crystal structure (5-7). The Born criterion is based on a rigidity catastrophe due to the vanishing elastic shear modulus. It has been demonstrated by molecular-dynamics (MD) simulations that the vibrational and elastic shear instability criteria coincide in determining the melting onset of an infinite Lennard-Jones crystal (8, 9). Apart from Lindemann's and Born's approaches to melting, a number of theories (10– 16) focused on the role of point (vacancies and interstitials) (11–13, 16) and line defects (dislocations) (14, 15). Surface-free, computer-generated systems which undergo a homogeneous melting process do not resemble real crystalline materials, which are spatially limited and often polycrystalline. It is well established that planar lattice defects, in particular surfaces and interfaces, play an essential role in crystal melting (17–25) giving rise to highly localized initiation or nucleation of the liquid. The presence of free surfaces and grain boundaries (GBs) are the main factors limiting superheating of real crystals and reaching the homogeneous melting regime (26, 27). It has been experimentally demonstrated that when surface melting is inhibited, metal crystals can be superheated appreciably above their equilibrium melting point $T_m$ (28). On the other hand, the presence of free surfaces (29) and GBs (22, 25, 30–32) can reduce melting onset temperatures below $T_m$. This phenomenon, predicted theoretically in the works of Frenkel (17), is known as premelting (19).

To achieve highly non-equilibrium states like strong crystal superheating, the timescale of the lattice heating must be shorter than that of its melting. It has been demonstrated that strong optical excitation of matter with femto- to picosecond laser pulses can result in extremely high heating rates, typically of the order of $10^{14}$ - $10^{15}$ K/s. The ultrafast dynamics of non-equilibrium melting requires a suitable probe to capture the complex, transient states of the photoexcited material. Several time-resolved methods have been used to study ultrafast melting of metals: optical reflectivity (33), X-ray absorption spectroscopy (34), diffraction of electrons (32, 35– 38) and X-rays (39– 41). Over the past few years, the electron and X-ray time-resolved diffraction techniques provided important information on the mechanisms of non-equilibrium melting. In particular, Mo *et al.* (41) employed ultrafast electron diffraction to visualize a transition between heterogeneous and homogeneous melting regimes in laser-excited Au. More recent work by Assefa *et al.* (*39*) reports the observation of a transient state during which the melting front propagates from the grain boundaries into the grain bulk.

Most of the previous studies of optically-driven melting were focused on weak electron-phonon coupling systems, such as Au (*21*, *35*, *39-42*). In those systems, the melting rate is limited by a slow energy transfer between the hot electrons and the atoms. The existing studies on systems with stronger electron-phonon coupling, such as tungsten (*32*), aluminum (*37*), or platinum (*43*), are either concentrated on the role of radiation-induced lattice defects and cover the low energy-density regime, where premelting rather than superheating is observed (*32*) or are not conclusive regarding the melting mechanism (*37*, *43*).

It remains an open question whether, in the case of strong coupling and fast equilibration between the excited electrons and the lattice, the melting mechanism changes or remains similar to that operating in the weak electron-phonon coupling systems. In particular, it is unclear whether at strong excitations the timescale of lattice heating can be distinctly shorter than that of melting. In such a case, the crystal would transiently undergo a significant overheating, and the actual rate of the crystal-liquid transition would be limited by the microscopic melting dynamics, not the lattice heating mechanism. From the experimental point of view, reaching this high energy-density regime would be manifested by saturation of the observed melting rate, with the maximum value characteristic of the transition mechanism.

The current work is focused on Pd, for which the electron-phonon coupling parameter was reported to be over five times higher than in Au (*44*). We investigate the ultrafast melting of thin polycrystalline Pd films by an optical laser pump – X-ray free-electron laser probe technique and large-scale two-temperature model molecular-dynamics simulations. Bridging the timescales of the experiment and the simulation allows us to formulate a realistic microscopic picture of the solid-liquid transition in an optically excited polycrystalline film and identify the melting mechanisms of metals with strong electron-phonon coupling.

## Experimental

Preparation and characterization of samples

The sample consisted of multiple X-ray transparent windows etched in a silicon wafer frame and arranged in a regular array (manufactured by Silson Ltd.). One frame contained ~2000 square windows, each of 300 x 300 μm size. The windows consisted of 300 nm thick $Si_3N_4$ membranes with a 28(2) nm thick Pd layer deposited on them by means of DC magnetron sputtering in an Ar atmosphere. The metal layer was capped with ~300 nm reactively sputtered $SiO_2$ (see Supplementary Material, Figure S1). The thickness of the deposited films was determined with X-ray reflectometry (XRR) measurements of witness samples (super-polished Si wafer co-coated during the same deposition run as the actual sample). It was further confirmed by transmission electron microscopy (TEM) measurements of the sample cross-sections. The TEM measurements also enabled the determination of the structure of each layer - the cap and the substrates were amorphous, while the Pd layer was polycrystalline, with crystallite sizes in the range of 5-30 nm. The XRR data, sensitive to the spatially averaged thickness of the layers, indicate an additional ~2 nm thick interface layer. However, the data do not allow distinguishing between a layer of different chemical composition (e.g., an oxide) or interface roughness (see Supplementary Material, Figure S2 for cross-section TEM micrographs of the films). The optical properties (reflectivity and transmission coefficients) of the samples (at the window positions) were measured with the same laser configuration as used for sample excitation.

Pump

The measurements were performed at the Femtosecond X-ray Experiments (FXE) instrument of the European XFEL (Germany) (*45*) in a set-up schematically presented in Figure S3 (Supplementary Material). 0.85 ps (FWHM) long laser pulses at 515 nm wavelength focused at almost normal incidence (3 deg incidence angle) on the sample in an approx. 105 µm diameter spot (square root of the "effective area" measured employing a method described in (*46*) were used to induce melting in the metallic film. The magnitude of the temperature rise of the metallic film was controlled by changing the pump pulse energy within the range of 0 to 40 µJ (corresponding to deposited energy densities from 0 to 3.1 MJ/kg, assuming the optical properties do not change during irradiation). The sample was irradiated at ~1 Hz repetition rate in a single-pulse mode. After each exposure to a single X-ray pulse, the sample was exchanged by moving to a "fresh" window on the sample array.

Probe

Femtosecond (pulse duration of 100 fs) X-ray pulses at 9.38 keV photon energy probed the central part (~10 µm diameter) of the laser-excited area. The delay time between pump and probe beams was controlled in the range from -0.1 ps (i.e. the X-ray probe pulse arrives before the optical pump pulse) to +60 ps. The scattered radiation was recorded with the Large Pixel Detector (LPD) available at the FXE instrument. It was placed off-center to monitor a momentum-transfer range up to $q \approx 4.5$ Å$^{-1}$ (limited by the diameter of the exit flange of the vacuum chamber). With up to $10^{11}$ photons (approx. 1 mJ total energy) per pulse at European XFEL, even a single X-ray pulse allows transient scattering patterns to be recorded with signal-to-noise ratio sufficient for data analysis without averaging, although the high X-ray fluence leads to subsequent sample modifications ("diffraction before destruction" (*47*)).

After necessary corrections, the 2D diffraction images were azimuthally integrated to obtain 1D intensity vs. momentum transfer *I(q)* patterns.

TTM-MD simulations

The time-resolved diffraction experiments are complemented by large-scale molecular-dynamics (MD) simulations coupled to a two-temperature model (TTM) description of the electron dynamics and the electron-lattice energy transfer, realized with our in-house computer code. Interaction of the ultrashort optical pump laser pulses with the Pd film establishes a strongly non-equilibrium situation with electron and ion subsystems having significantly different temperatures. To properly address such a two-temperature state of Pd, its thermal properties, namely the electron heat capacity, $C_e(T_e)$, and the electron-phonon coupling factor, $G(T_e, T_i)$, were obtained using separate calculations based on the tight-binding MD method (*44, 48*). Here, $T_e$ and $T_i$ are the electron and ion temperatures, respectively. The electron thermal conductivity of Pd in the two-temperature state is not available, to our knowledge. Therefore, we used an experimental value of 122 W/m/K measured for solid Pd at 1800 K (*49*).

The interatomic interactions are taken into account in the MD simulations with a newly developed interatomic potential for Pd atoms (details are provided in the Supplementary Material).

The TTM-MD simulation setup consists of a 30 nm thick Pd polycrystalline film sandwiched between two layers of an effective amorphous material representing the $Si_3N_4$ substrate and the $SiO_2$ cap. The polycrystalline structure of Pd is modeled in the first step using the Voronoi tessellation method (*21*). The resulting hexagonal (in-plane) and columnar (out-of-plane) grains are then melted and recrystallized to obtain a more polycrystalline structure, better corresponding to the one studied experimentally. The effective material of substrate and cap is modeled by a simple Morse potential (*50*) consisting of effective atoms with the mass (and three parameters of the potential) adjusted to produce the $Si_3N_4$ substrate mechanical properties, namely density and speed of sound. To better represent the experiment, the simulated amorphous structure was prepared by temporally decreasing the depth of the Morse potential in order to have the melting temperature lower than that of in Pd, and then melting and quenching at room temperature. The resulting structure of the effective material is amorphous except for a narrow region at the interface with the polycrystalline Pd, where recrystallization occurred. The final depth of the Morse potential is chosen so that the melting point of the effective material is higher than that of Pd, as expected. Its thermal conductivity is found to be close to that of $Si_3N_4$. The thickness of the effective layers is adjusted to 320 nm and 487 nm for substrate and cap, respectively. Such a choice represents realistic behavior of the pressure waves induced by laser heating, as will be explained in more detail below. We do not aim to quantitatively recreate the properties of both $Si_3N_4$ substrate and

SiO$_2$ cap to the full extent, but rather model these materials with effective layers that act as a realistic heat sink and a medium for realistic pressure dynamics in the whole system.

The in-plane cross-section of our TTM-MD simulation box is 23 × 20 nm$^2$, and periodic boundary conditions are applied in these directions. With such a simulation geometry, we assume that the in-depth (out-of-plane) evolution of the laser-irradiated sample is dominant on the timescales we consider. This assumption is justified since the laser spot size is much larger than the thickness of the Pd layer that absorbs the laser pulse. The lateral (in-plane) heat diffusion that may play a role on long timescale (microseconds and longer) is not taken into account.

Heat dissipation from the laser-melted Pd into the substrate and cap proceeds via phonon thermal conductivity accounted with the MD method. A direct coupling of hot electrons in Pd to substrate/cap phonons at the interfaces, which usually becomes relevant for small film thicknesses (*51*, *52*), is not considered here. Therefore, the heat dissipation from Pd in our model may be underestimated.

In a metal, the laser energy is absorbed by the conduction band electrons. The thickness of the Pd layer is comparable to the pump laser penetration depth. Therefore, uniform distribution of the absorbed energy is established fast, approximately within the first picosecond via electronic heat diffusion. Hence, the TTM-MD simulations start at an elevated uniform electron temperature profile, corresponding to a particular absorbed energy density, whereas the metal ions are at room temperature. Fast energy exchange between hot electrons and cold ions through electron-phonon coupling results in a rapid increase of the ion temperature, resulting in a simultaneous increase of thermal pressure. This process initiates rarefaction waves into the Pd-layer (and corresponding compression waves into the substrate and the cap) that start at the metal-substrate and metal-cap interfaces. The complex dynamics of pressure waves is captured in our TTM-MD simulations. In the real sample, the Si$_3$N$_4$ and SiO$_2$ layers have equal thicknesses, but the speed of sound is different. Therefore, the pressure waves traveling in these materials return to Pd at different times after reflecting from the front/rear sides of the sample. To mimic this, we model our substrate and cap layers having different thicknesses, as stated above.

With our TTM-MD simulations, we were able to trace the evolution of the laser-irradiated sample over the entire timescale from laser absorption to melting. The spatiotemporal scale of our simulations corresponds to the experimental conditions, which enables a one-to-one comparison with the experimental data. For this purpose, the MD configurations were analyzed and visualized using the OVITO package (*53*). From these configurations, X-ray diffraction (XRD) patterns were calculated for the incident X-ray photon energy of 9.38 keV (corresponding to a wavelength of 1.32 Å) using a computational method developed by Coleman (*54*) and implemented into the LAMMPS code (*55*).

## Results

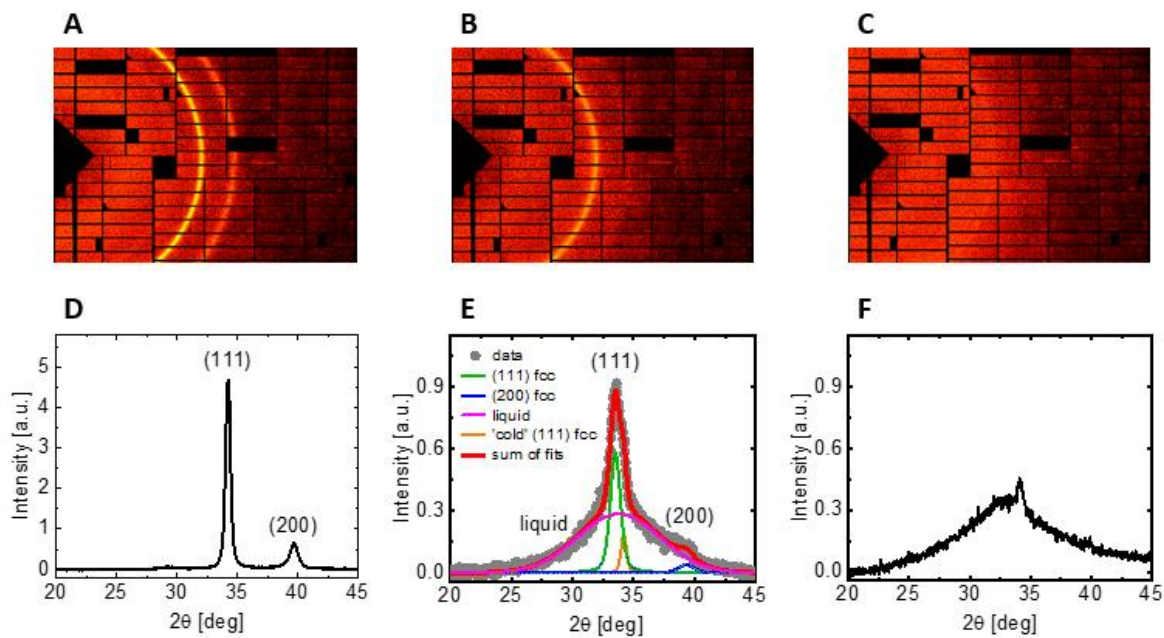

**Figure 1. XRD snapshots representative of different stages of the melting transition.** XRD patterns of a crystalline (**A**), partially molten (**B**), and fully molten Pd film (**C**) and the corresponding azimuthally-integrated, one-dimensional patterns (**D**, **E**, **F**). The deconvolution of the diffraction patterns into contributions from the crystalline and the liquid phase is shown in (E).

X-ray diffraction

In the obtained experimental data, the melting transition was manifested by variations of the XRD patterns acquired for different excitation energy densities and different pump-probe delay times. Figure 1 shows XRD snapshots (2D detector images and azimuthally integrated 1D patterns) representative of different stages of the solid-liquid transition. As confirmed by the presence of fine Debye-Scherrer rings assigned to the (111) and (200) Bragg reflections of face-centered cubic (fcc) Pd, the as-deposited film (Fig. 1A and 1D) is fully crystalline. A weak peak located at $2\theta = 29$ deg is present only in the "as-deposited" state and can be ascribed to a negligibly small volume fraction of an unidentified crystalline phase observed by the TEM at the interface between the metallic film and the substrate. The uniform character of the rings with no distinct spots reveals a nanocrystalline character of the film with no apparent texture or preferred orientation of the crystalline grains, in agreement with the TEM analysis of the as-deposited films (see Supplementary Material). In the partially molten state, the XRD pattern of the metallic film consists of fine rings and a diffuse "halo" originating from the liquid phase (Fig. 1B and 1E). The fully molten state (Fig. 1C and 1F) is manifested by the lack of Debye-Scherrer rings and the presence of the broad liquid peak only (the weak persistent fcc (111) Bragg reflection visible in Fig. 1E and 1F is due to scattering from non-excited

("cold") material resulting from weak, but large-diameter tails of the focused X-ray beam). Figure 1E presents the results of the numerical deconvolution of the diffraction pattern of the partially molten film into contributions from crystalline and liquid material. As shown in the plot, the experimental data can be accurately reproduced by a sum of a broad Gaussian peak representing the liquid peak and narrow Pseudo-Voigt peaks corresponding to the (111) and (200) Bragg reflections of fcc Pd. Deconvolution of a 1D diffraction pattern thus yields values of the area, width, and position of each peak – all quantities related to the physical state of the probed film. Since the "cold" fcc (111) peak remained constant throughout the melting transition it was treated as a background and subtracted from the data, as discussed in the following.

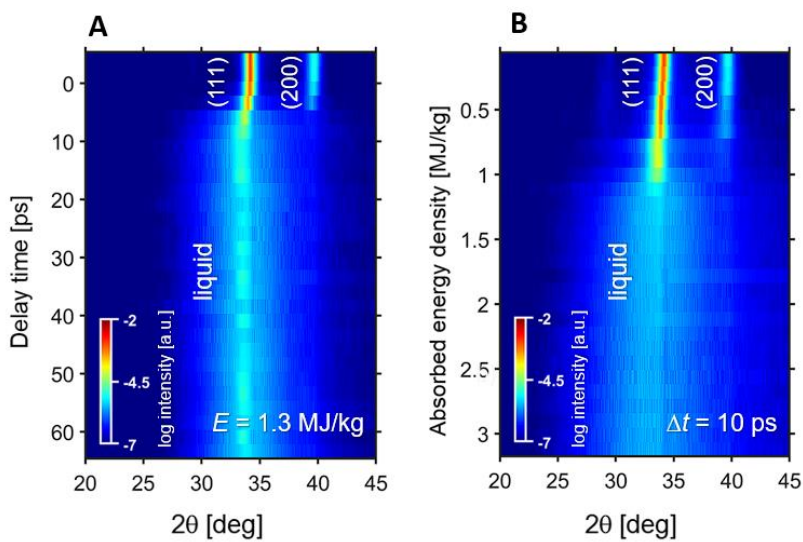

**Figure 2. Delay time and absorbed energy density dependence of the integrated XRD patterns.** False-color plot representation of the temporal evolution of the diffraction pattern at $E$ = 1.3 MJ/kg (**A**) and its $E$-dependence at a pump-probe delay of 10 ps (**B**). Bragg peaks of Pd fcc and the broad liquid peak are indexed.

The two key parameters controlled in the experiment were the deposited pump energy density per unit mass ($E$) and the pump-probe delay time ($\Delta t$). Therefore, we will use in the following two different representations of the experimental data, namely $\Delta t$-dependences of a selected quantity for a fixed $E$ ("$\Delta t$-scan"), and $E$-dependences for a fixed $\Delta t$ ("$E$-scan"). Examples are depicted in Fig. 2, which show the variation of the diffraction signal in a false-color representation. Fig. 2A shows a $\Delta t$-scan at $E$ = 1.3 MJ/kg. While no variation of the diffraction pattern is observed for negative delay times (X-rays arriving before the pump pulse), for positive delays an initial shift of Bragg peaks towards lower scattering angles and a decrease of their intensity is observed. The shift is attributed to the thermal expansion of the crystalline lattice, while the decrease in the peak intensity is a combined effect of atomic thermal vibrations (Debye-Waller effect) and the progress of melting. Along with the progress of melting, a broad liquid signal develops, and about 10 ps after the excitation the film is fully molten. Figure 2B presents a similar false-color representation of the XRD patterns from an $E$-scan at $\Delta t = 10$ ps. This allows us to determine the energy density required for complete melting of the film at $\Delta t = 10$ ps to $E \approx 1.3$ MJ/kg.

For a detailed analysis of the melting transition, we determine the integrated diffraction signal $A_C$ of the fcc (111) Bragg peak (normalized to its value without optical excitation i.e. at negative delay times), and $A_L$ of the liquid peak (normalized to the value of the fully molten sample). In addition, to gain information on the variations of the crystalline microstructure on irradiation, we traced the peak width $\Delta\Theta_{111}$ (full width at half maximum - FWHM) of the (111) peak. The melting dynamics is thus expressed in terms of $A_C$, $A_L$, and $\Delta\Theta_{111}$ as a function of $\Delta t$ and $E$. We note that the intensity of the fcc (200) peak was found to be too low (for example, see Fig. 2E) for a reliable peak position and profile analysis.

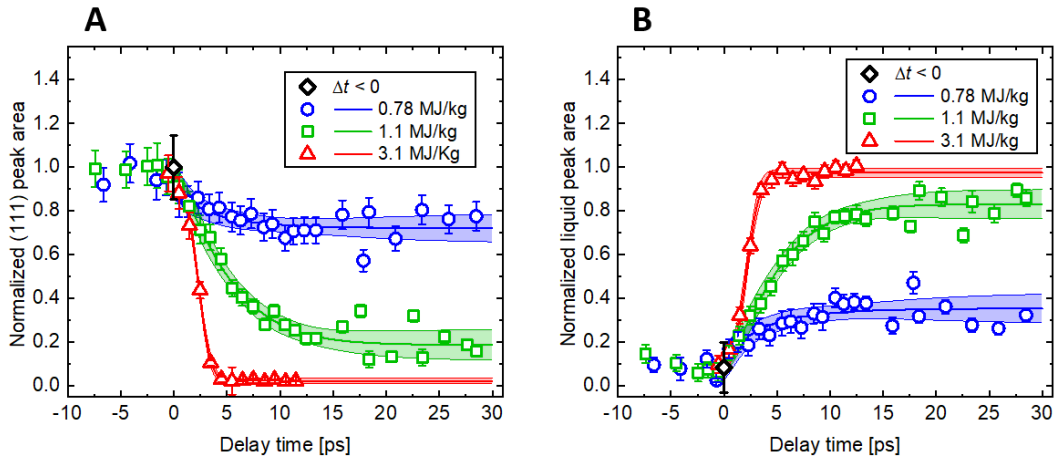

**Figure 3. Delay time dependence of the crystalline and liquid contributions to the diffracted signal.** Delay time-scans of the normalized fcc (111) Bragg peak area (**A**) and liquid peak area (**B**). Symbols represent experimental data points; lines are best fits using an empirical function (see text). The symbol "$\Delta t < 0$" represents the average peak area for the negative delay times, and its error bars correspond to three standard deviations. Shaded areas are confidence bands of the fits.

Figure 3 shows $\Delta t$-scans of $A_C$ and $A_L$ for three selected values of $E$ (0.78 MJ/kg, 1.1 MJ/kg, and 3.1 MJ/kg). The data point denoted as "$\Delta t < 0$" refers to the average value for the negative delay times (non-irradiated sample). As seen in Fig. 3A, after the optical excitation, an immediate drop of the crystalline signal occurs, with the magnitude and the rate of the drop proportional to the deposited energy density. For $E = 0.78$ MJ/kg, $A_C$ drops to approx. 0.75, for 1.1 MJ/kg it reaches 0.2, and for 3.1 MJ/kg – 0.02. The decrease of the crystalline signal is accompanied by a growth of the liquid contribution to the diffracted intensity (Fig. 3B). To quantify the temporal variations of $A_C$ and $A_L$, we fitted the data plotted in Fig. 3 with an empirical function $A_i(\Delta t) = A_i^0 + \left(A_i^0 - A_i^{\mathrm{inf}}\right) \cdot \left(e^{-(\Delta t/\tau_i)^{n_i}} - 1\right)$ with $i = C, L$ representing the crystalline and the liquid contribution, respectively. For $\Delta t = 0$, the value of the function is

$A_i^0$ (set to 1 for fitting of $A_C$ and 0 in the case of $A_L$) and subsequently decreases approaching $A_i^{\text{inf}}$ in the infinite time limit, at a rate dependent on the time constant $\tau_i$ and a dimensionless exponent $n_i$.

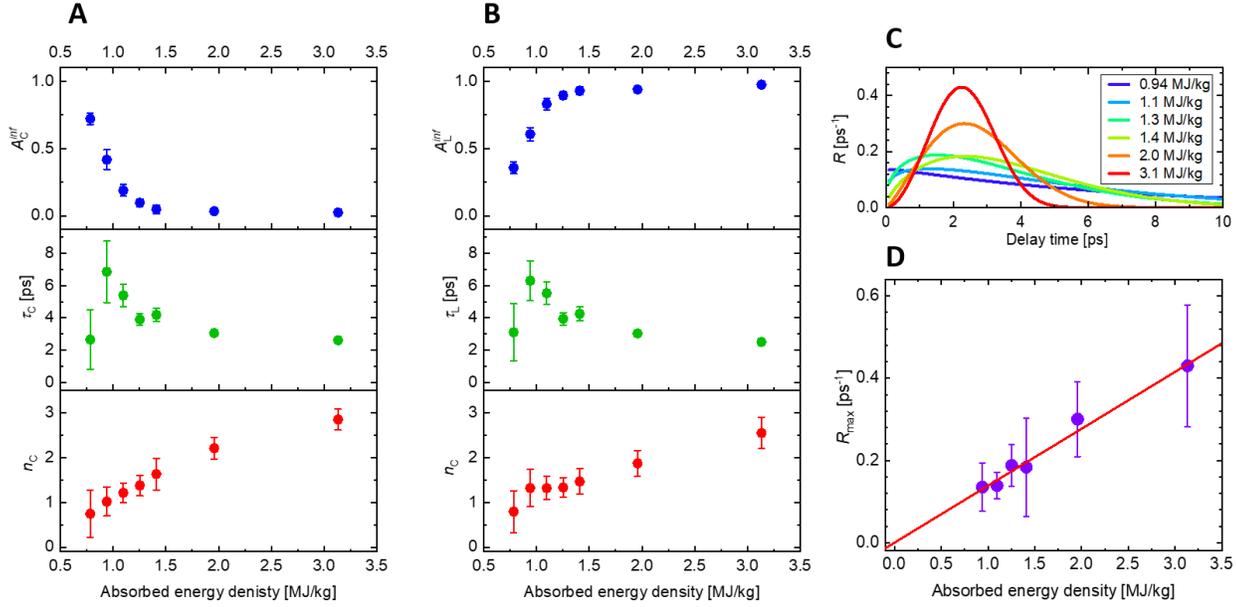

**Figure 4. Fitting parameters of the temporal variation of the crystalline and liquid peaks.** Best-fit values of the integrated diffraction signal $A_i$, time constant $\tau_i$, a dimensionless exponent $n_i$ as a function of the absorbed energy density for (**A**) crystalline peak, (**B**) liquid peak, (**C**) delay time scan of parameter $R$ illustrated normalized rate of the (111) Bragg peak intensity, (**D**) its maximum value $R_{\text{max}}$ as a function of absorbed energy density.

Figure 4 presents the best-fit parameters obtained for the crystalline (Fig. 4A) and the liquid (Fig. 4B) peak plotted as a function of the absorbed energy density. As expected, the saturation values $A_C^{\text{inf}}$ and $A_L^{\text{inf}}$ sum up to unity, indicating that the decrease of the crystalline signal is compensated by an increase of the liquid contribution. The kinetic parameters $\tau_i$ and $n_i$ exhibit similar $E$-dependence for the crystalline and the liquid phase, reflecting the increasing rate in the variation of $A_C$ and $A_L$ with increasing density of absorbed energy. To better visualize the acceleration, we introduce a parameter $R = \left|\left(\frac{dA_C}{d\Delta t}\right)/\Delta A_C\right|$, where $\Delta A_C = A_C^0 - A_C$. The interpretation of the parameter is a normalized rate of decrease of the (111) Bragg peak intensity. Figure 4C presents the $\Delta t$-dependence of $R$; its maximum value, $R_{\text{max}}$ as a function of $E$ is shown in Figure 4D. The correlation between $R_{\text{max}}$ and $E$ suggests that the rate of decrease of the crystalline contribution to the XRD pattern approximately scales linearly with the absorbed energy density, without any apparent signs of saturation in the studied range.

The $E$-dependent temporal evolution of $A_C$ and $A_L$ is presented in Figure 5 in the form of $E$-scans for three selected values of $\Delta t$: 5 ps, 10 ps, and 30 ps, together with data points corresponding to $A_C^{\text{inf}}$ and $\Delta t < 0$. According to Fig. 5A, for the energy density in the rage from 0 to 0.8 MJ/kg the decrease of $A_C$ is identical for the three delay times. In this range, $A_C$ decreases monotonically, and the data points overlap within the error bars. Around 0.8 MJ/kg, the points for 5 ps start to deviate from those for 10 ps and 30 ps, and those representing $A_C^{\text{inf}}$. Around 2.0 MJ/kg the points converge again. The evolution pictured in Fig. 5A indicates the presence of two processes affecting the intensity of the Bragg peaks. The first one, occurring on a time scale shorter than 5 ps, is responsible for the reduction of intensity by up to ~25% at 0.8 MJ/kg. The second one sets in around 0.8 MJ/kg. For low $E$ it proceeds on a time scale shorter than 10 ps, and leads to a further decrease of $A_C$ down to ~ 0.02 (corresponding to the background noise level and considered further as a zero). It accelerates with increasing $E$ and eventually is completed in less than 5 ps when excitation is stronger than 2.0 MJ/kg.

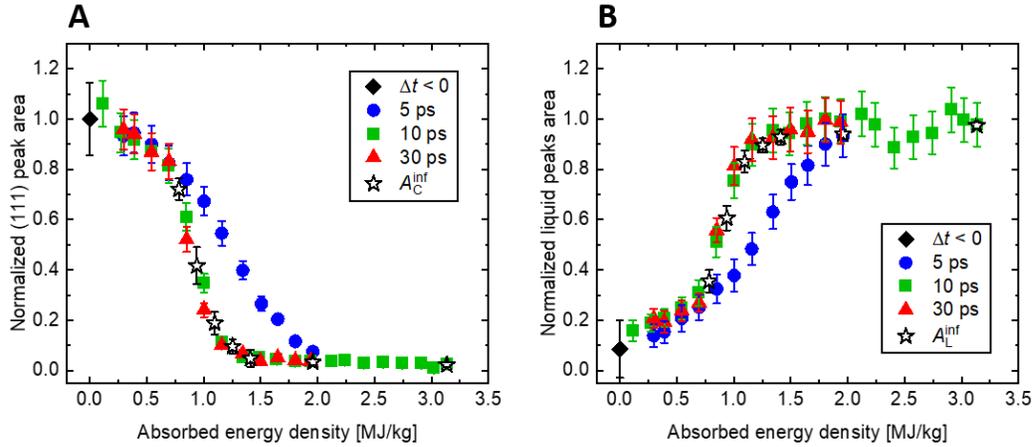

**Figure 5. Energy density dependence of the crystalline and the liquid contributions to the diffracted signal.** Normalized areas of the (111) Bragg (**A**) and the liquid peak (**B**) as a function of absorbed energy density plotted for selected values of pump-probe delay time. The symbol "$\Delta t < 0$" represents the average peak area for the negative delay times, and its error bars correspond to three standard deviations. $A_C^{\text{inf}}$ and $A_L^{\text{inf}}$ denote saturation values of the fits with an empirical function (see text).

According to Fig. 5B, the maximum fraction of the liquid phase increases with $E$ up to ~ 1.4 MJ/kg when the data points for 10 ps, 30 ps and $A_C^{\text{inf}}$ become independent of $E$ and overlap within error bars at zero. The energy density of 1.4 MJ/kg thus marks the upper limit of partial melting and is followed by the regime of complete melting. Above 1.4 MJ/kg, the decreasing vertical gap between the data points for different delay times reflects the increasing melting rate with increasing deposited energy density $E$. The gap closes at 2.0 MJ/kg, indicating that at this strong excitation levels the melting process is completed in less than 5 ps after the excitation.

Figure 6A presents $\Delta t$-scans of $\Delta\Theta_{111}$ for the absorbed energy densities of 0.78 MJ/kg, 1.3 MJ/kg, and 3.1 MJ/kg, together with the "$\Delta t < 0$" data point. The mean crystalline grain size derived from the peak width of the non-excited film is estimated using the Scherrer formula (*56*) as ~31 nm, which is close to the upper limit of the crystal size range observed by TEM (Figure S2). Due to the significant uncertainty in the width determination for low peak amplitudes, we omit the fitting results for $\Delta\Theta_{111}$ at amplitudes below 5% of the maximum value. In contrast to $A_C$, the width $\Delta\Theta_{111}$ does not start to vary immediately after the excitation. For 0.78 MJ/kg and 1.3 MJ/kg, an increase in the peak width is preceded by a plateau regime of approximately 3 ps. After the sharp rise lasting approximately 10 ps, $\Delta\Theta_{111}$ reaches a maximum and subsequently decreases but does not reach its initial value. For 3.1 MJ/kg, the rapid decrease of $A_C$ (see Fig. 3A) excludes a reliable determination of the peak width beyond the initial transient of 3 ps. Therefore, only data points corresponding to the initial plateau regime are shown. More extensive information on the variation of $\Delta\Theta_{111}$ can be deduced from $E$-scans measured at different pump-probe time delays, as shown in Figure 6B. For energy densities below 0.8 MJ/kg, the width does not change during the initial 5 ps after the excitation. At higher energy densities, the increase of $\Delta\Theta_{111}$ becomes more rapid, and its maximum corresponds to data points for 10 ps, in agreement with $\Delta t$-scans in Fig. 6A.

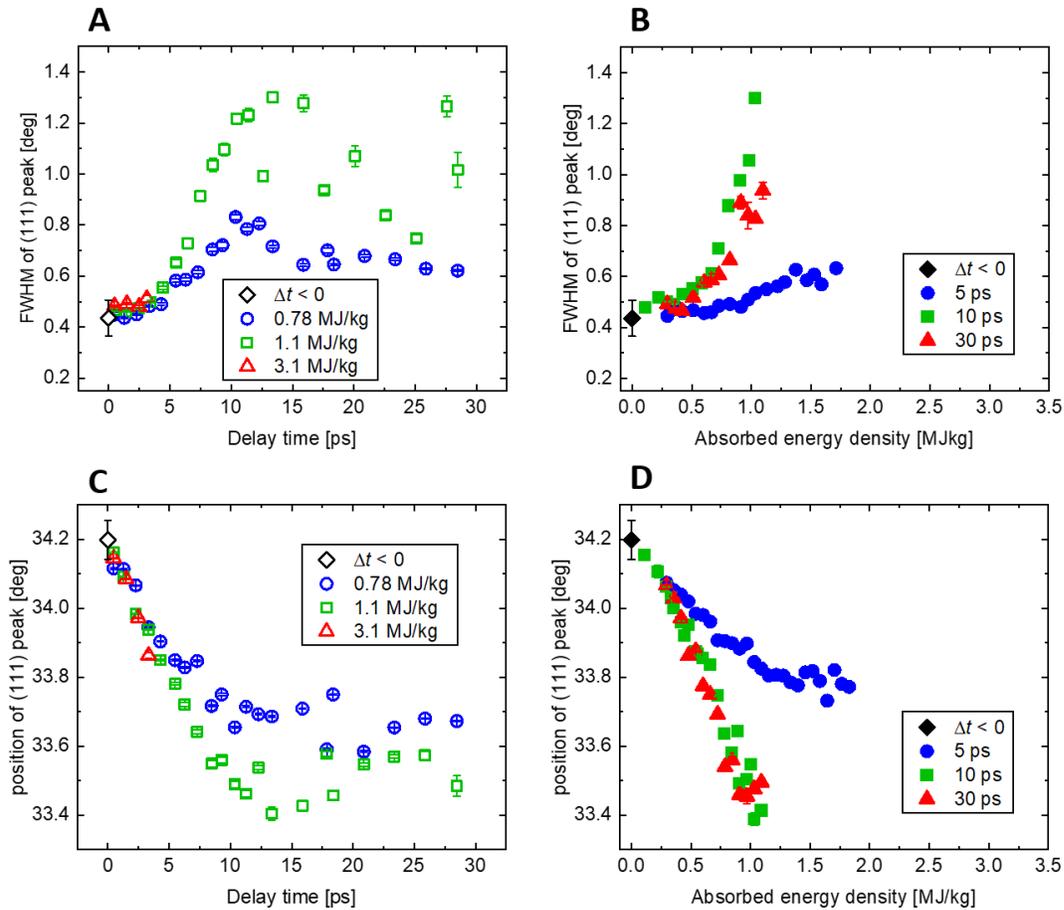

**Figure 6. Delay time and absorbed energy density dependence of the position and the width of the fcc Pd Bragg peak.** Delay time (**A**, **C**) and absorbed energy density scans (**B**, **D**) of Pd fcc (111) Bragg peak width (**A**, **B**) and position (**C**, **D**). The data points are shown in the $\Delta t$-range, where the peak parameters can be reliably determined.

Apart from the peak width, we examined the variation of the (111) peak position to find a correlation with its width. Figures 6C and 6D show the $\Delta t$ and $E$-scans of the peak position. According to Fig. 6C, the (111) peak progressively shifts towards lower scattering angles for about 10 ps after the excitation and remains roughly constant during subsequent 20 ps. This behavior is confirmed by the $E$-scan shown in Fig. 6D. The maximum shift by approximately 0.7 deg towards lower angles indicates an expansion of the crystalline lattice by up to 2% in the [111] crystallographic direction (assuming 3D-expansion). Such a process could be driven by the thermally induced increase of pressure inside the metallic film (*57*).

Molecular-dynamics simulations

While the XRD patterns of the transient states of the metallic film provide essential information on its evolution from the crystalline to the liquid state, determining the atomic structures corresponding to those patterns is challenging, as discussed in the previous section. This fundamental problem arises because significantly different atomic arrangements can yield similar diffraction patterns – the non-uniqueness being a consequence of the inverse scattering problem (*58*). To reveal the complete structural information encoded in the transient XRD data, we have conducted extensive TTM-MD simulations of the melting process and used the simulated configurations to derive the corresponding XRD patterns. The unique feature of this approach is a one-to-one correspondence of the timescale and characteristic system size (the film thickness) between the simulation and the experiment, allowing a direct comparison of the experimental data with the theoretical predictions. We underline that due to intrinsic limitations of the MD simulations (accuracy of the empirical interatomic potential, arbitrary choice of the initial atomic configuration), a perfect quantitative agreement between experiment and simulation cannot be expected. Nevertheless, the MD simulations yield a reliable, physical model of the system during the transition from the (poly)crystalline to the liquid state.

An essential issue in matching experiment and simulations is the per-se unknown relation between the deposited energy density in the experiment (derived from the incident pump fluence and measured optical transmission/reflection coefficients of the film) and the theoretical initial energy density used in the TTM-MD simulation. In theoretical studies of optically excited metals, this ratio between the experimental and the theoretical energy density is often treated as a free parameter (*35*, *36*). The reason is the existence of multiple paths of energy loss (e.g., ballistic electron motion (*59*), coupling with the localized vibrational modes on the metal-substrate interface (*52*)), which depend on the details of the sample (its layout and thermal properties of materials involved) and which are challenging to incorporate in the simulation realistically.

Another vital aspect of the comparison between the experiment and the simulation is related to the limited size of the simulation box. While in the experiment, the size of the focused X-ray beam is approximately 10 µm, the dimension of the (in sample plane) cross-section of the simulation box is about 20 nm (Figure 7A). Considering equal thicknesses of the experimentally probed and the simulated Pd films (~30 nm), the ratio of the scattering volume in the simulation is more than five orders of magnitude smaller than that in the experiment. Using the relative orientation of the incident X-ray beam and the film as in the experiment (i.e. X-ray beam parallel to the film surface normal) for calculation of the XRD-patterns from the MD-simulated system, the resulting two-dimensional diffraction pattern consists of just a few diffraction spots originating from the very limited number of individual crystalline grains in the simulation cell satisfying Bragg's condition for this orientation. Such a diffraction pattern is practically useless for quantitative analysis, due to statistically insignificant amount of structural information resulting in irregular, spiky diffraction peaks. To overcome this difficulty, but taking also into account that it is totally unrealistic to perform a sufficient number of simulation runs for simulation cells with different grain structure (this would be actually equivalent of doing one run with a transverse size of the simulation cell comparable to the X-ray spot size in the experiment), we averaged 1800 diffraction patterns with a systematically varied direction of the X-ray beam relative to the simulation cell. While this makes the resultant patterns analyzable and equivalent to those obtained in a powder diffraction experiment, it does not fully correspond to the experimental situation with respect to strain induced changes. For low-order Bragg peaks with correspondingly small diffraction angles momentum transfer is predominantly perpendicular to the direction of the incident X-rays. As a consequence the peak position is substantially less sensitive to strain along the direction of the X-rays, compared to strain perpendicular to it. In the experiment, the X-ray probe is parallel to the film surface normal, and the experimental diffraction patterns are much less sensitive to strain out of the film plane than in-plane strain. Consequently, uniaxial strain along the surface normal will lead to a much smaller shift of such a low order Bragg peak than isotropic strain of the same magnitude (*57*, *60*). In comparison, due to the way the simulated diffraction patterns are derived as an average of patterns with random orientation of the X-ray beam to the surface normal of the simulation cell, the Bragg peak positions in the simulated patterns are equally sensitive to out-of-plane and in-plane strain. A similar difference can be expected for the variation of the peak width originating from non-uniform strain. Finally, it should be emphasized that the limited size of the MD-simulated polycrystalline Pd film constrains the crystalline grain size. While the experimentally determined mean grain size of the metallic films studied is 28 nm, the corresponding number for the simulated system is estimated as 10 nm. The consequences of this inconsistency are two-fold. Firstly, since the peak width is inversely proportional to the mean crystallite size (*61*), already the initial peak width of the unexcited material in the simulated XRD patterns will be larger than that observed in the experiment. Secondly, smaller crystallites in the simulated system size imply a higher content of the grain boundary regions, thus enhancing the expected contribution of heterogenous melting initiated at GBs. This difference will affect the details of the melting kinetics and, consequently, the transient changes of the diffraction patterns. Therefore, a full quantitative agreement between simulation and experimental results should not be expected. Nevertheless, as we will show below, our as-realistic-as-possible TTM-MD simulations and the derived XRD patterns are able to reproduce and explain the main trends observed in the experiment.

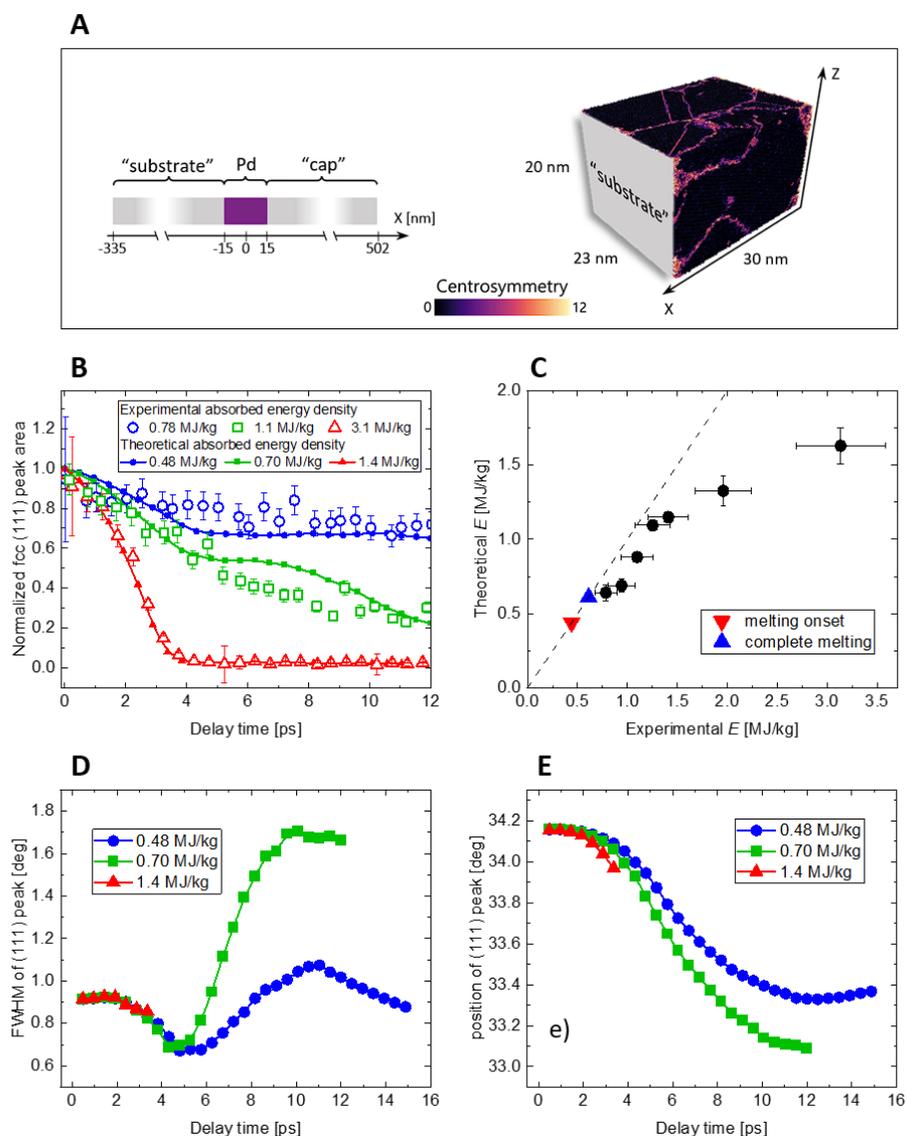

**Figure 7. Parameters of the XRD patterns calculated for TTM MD simulations of optically excited polycrystalline Pd film.** Layout of the TTM-MD setup and a polycrystalline structure of the simulated Pd film (**A**). Matching of the theoretical (full symbols + lines) and experimental (open symbols) temporal variations of the normalized area of Pd Bragg peaks is shown in (**B**), and (**C**) plots the estimated correlation function between the theoretical and the experimental absorbed energy densities. The dashed line in (**C**) represents a one-to-one correlation between the energy densities, and the triangles represent the melting onset and complete melting threshold calculated from electron and lattice heat capacities, melting temperature and melting enthalpy. Delay time scans of the Pd fcc (111) Bragg peak width (**D**) and position (**E**) derived from the MD-simulated atomic configurations. The data points in (**D**) and (**E**) are shown in the $\Delta t$-range, where the peak parameters can be reliably determined.

Bearing in mind the constraints in calculating XRD patterns from the simulation results, we have systematically matched the experimental and the simulated $\Delta t$-scans of $A_C$ to derive the correlation between the experimental and the theoretical energy densities. Figure 7B presents matched experimental and simulated $\Delta t$ - scans of $A_C$ for selected pairs of energy densities. A systematic matching of the different pairs of the $\Delta t$ - scans of $A_C$ (see Supplementary Material) yielded an empirical correlation between the theoretical and the experimental $E$ depicted in Figure 7C. While, as pointed out above, the form of the correlation is generally unknown, it is reasonable to expect that zeros of the actual and the assumed $E$ coincide. Extrapolation of the low-$E$ data points in Fig. 7C suggests that this simple condition is satisfied within the estimated error bars. Furthermore, according to the predictions in the available literature, the energy loss by the optically excited metal becomes more pronounced at high energy densities, reaching up to 80% at strong excitations (*36*). The effect is manifested in Fig. 7C by a decreasing slope of the plot. A combination of good agreement between the experimental and the simulated temporal evolution of $A_C$ and a reasonable correlation between the assumed and the actual energy densities is a strong argument supporting the relevance of the current TTM-MD simulations for the performed experiment.

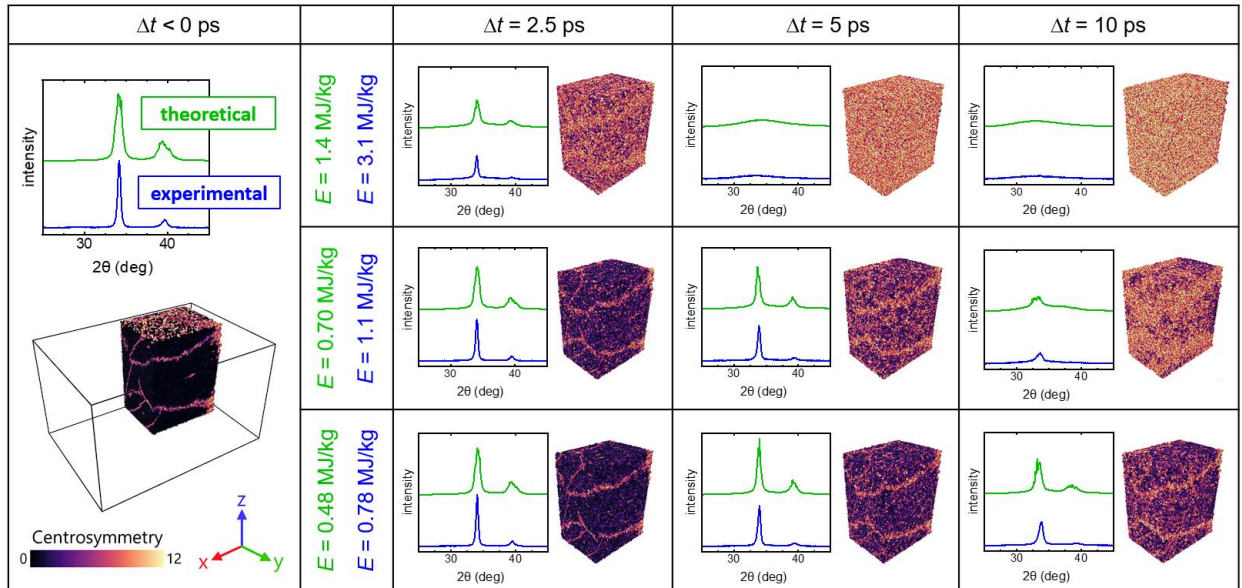

**Figure 8. Structural evolution of the laser-irradiated polycrystalline Pd film resulting from TTM MD simulations.** The film is oriented in the *yz* plane. The snapshots are shown for negative delay time, 2.5 ps, 5 ps, and 10 ps, and theoretical $E$ of 0.48 MJ/kg, 0.70 MJ/kg, and 1.4 MJ/kg. The theoretical patterns are matched with experimental ones measured at corresponding delay times and energy densities of 0.78 MJ/kg, 1.1 MJ/kg, and 3.1 MJ/kg. The color of atoms corresponds to the value of their centrosymmetry parameter. The intensity scale of the experimental and the theoretical XRD patterns does not vary across the plots.

Figures 7D and E show simulated dependences on $\Delta t$ of the Pd fcc (111) peak position (E) and width (D). Apart from the dip around 4-5 ps, the width rises sharply, reaches a maximum at around 10 ps, and then decreases. Regardless of the dip (presumably related to transient relaxation of the residual stress developed on recrystallization of the film during sample preparation stage), the simulated variation of the peak width resembles the behavior observed in the experiment (compare Fig. 6A). According to Fig. 7E, our model also reasonably reproduces the temporal variation of the peak position (compare with Fig. 6C) and its correlation with the peak width (the position of $\Delta\Theta_{111}$ maximum coincides with a knee of the peak position plot). Just as described above, both the values of $\Delta\Theta_{111}$ and the magnitude of the peak shift obtained from the simulated XRD patterns are higher than those observed experimentally.

The qualitative and partially quantitative agreement between the experimental and simulated $\Delta t$ and $E$-dependences of integrated diffraction efficiency, width, and position of the (111) Bragg peak have two important implications. Firstly, it reinforces the interpretation of the experimental data by avoiding the previously mentioned problem of the intrinsic non-uniqueness of the diffraction pattern. Secondly, it justifies further steps of MD data treatment, namely a detailed analysis of atomic configurations corresponding to transient states during laser melting. Figure 8 pictures the $E$-dependent temporal evolution of the optically excited polycrystalline Pd film. The snapshots present the atomic configurations of the cross-section of the MD system with atoms colored according to the centrosymmetry parameter (CSP). The interpretation of the CSP, introduced originally by Kelchner *et al.* to distinguish between different types of crystal defects (*62*), is the deviation from the centrosymmetry. The CSP (expressed in the units of Å$^2$) is zero for an atom in a perfect fcc lattice, approx. 2 for atoms near partial dislocations, reaches about 8 for atoms in an intrinsic stacking fault, and exceeds 20 for surface atoms. Besides the film's structural snapshots, Figure 8 shows the corresponding experimental and simulated XRD patterns. At $\Delta t < 0$, apart from narrow GB regions, the film's structure is highly ordered (low CSP), and the resultant XRD patterns exhibit relatively sharp Bragg peaks of the fcc lattice. At the lowest energy density (the experimental and the theoretical energy densities of 0.78 MJ/kg and 0.48 MJ/kg, respectively, just below the melting onset), the CSP globally increases with time. The effect can be attributed to the increasing thermal disorder of the atomic lattice and the growing concentration of lattice defects. As a result of atomic thermal vibrations, an apparent decrease of the Bragg peaks' intensities is observed. With rising energy density (experimental $E = 1.1$ MJ/kg, theoretical $E = 0.7$ MJ/kg) the thickness of the disordered GB regions noticeably increases with $\Delta t$. The effect is interpreted as heterogeneous melting initiated at the GBs and propagation of the transformation front into the crystals' centers. The propagation velocity was estimated from the temporal variation of the high-CSP GB regions to be approximately 400 m/s. At this energy density, the system state seen by the XRD after 10 ps corresponds to a broad liquid peak with traces of the residual (111) Bragg reflection. At the highest energy density (experimental $E = 3.1$ MJ/kg, theoretical $E = 1.4$ MJ/kg), the homogeneous melting mechanism dominates the transition. The homogeneous character of melting is manifested by an increased uniformity of the spatial distribution of the high-CSP atoms. In the partially molten (as seen by the XRD) state at $\Delta t = 2.5$ ps, the contrast between the GBs and the bulk is barely visible. With homogeneous melting being the dominant melting mechanism, the film is entirely molten in less than 5 ps after the excitation.

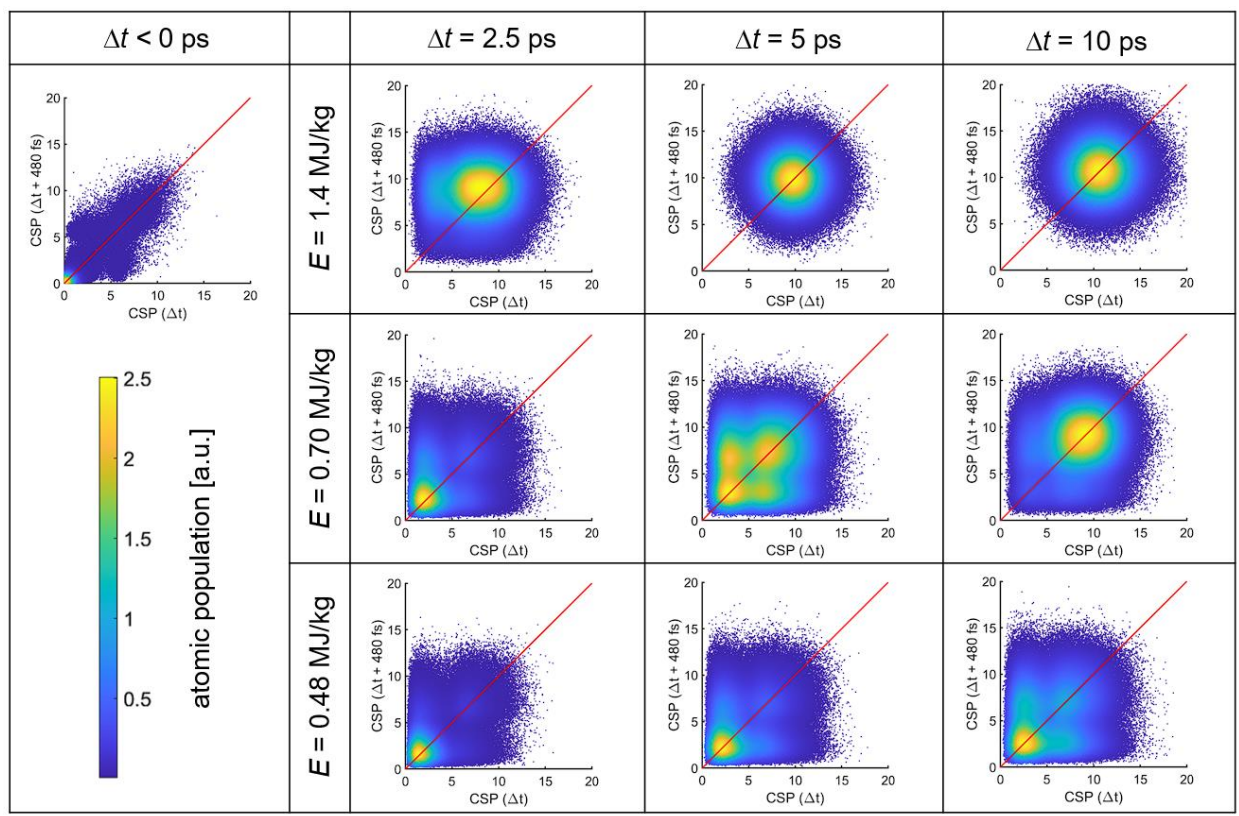

**Figure 9. Variation of the atomic centrosymmetry parameter in polycrystalline Pd following the optical excitation**. 2D histograms presenting the evolution of the CSP for selected $\Delta t$ and $E$ values matching those in Figure 8. The time of the vertical CSP axis is shifted by +480 fs with respect to the horizontal one. The red diagonal lines represent $CSP(\Delta t) = CSP(\Delta t + 480\ fs)$.

The CSP-based visualization of the melting process presented in Fig. 8 shows the departure of atoms from the centrosymmetry and thus helps resolve the spatial distribution of the crystalline and liquid phases during melting. However, in the cross-section picture, only the surface atoms are visible, and thus the overall information on the CSP distribution is missing. To follow in detail the behavior of the CSP during melting, we analyzed the MD configurations to extract the global variation of the CSP. Figure 9 presents the evolution of the CSP as a 2D histogram with time corresponding to the vertical axis shifted by 0.48 ps with respect to the horizontal one. The time shift value was chosen to exceed the dominant period of atomic oscillations in Pd crystal (~200 fs, as estimated from high-temperature phonon density of states of Pd (*63*)) and, at the same time, to remain well below the shortest observed melting time (~4 ps, according to the current data). In this representation, an atom's centrosymmetry change consists of a series of steps in the vertical direction (the variation of its CSP during the preceding 0.48 ps) taken at different positions on the horizontal axis (for different initial values of the CSP). The position at the plot diagonal marks an invariant centrosymmetry, while the off-diagonal position represents a transient increase (points above the diagonal) or decrease (points below the diagonal) of the CSP. For the evolution pictured in Fig. 9, the vertical step size is sufficiently large to exceed CSP variations related to thermal atomic oscillations. On the other hand, the 2D histograms plotted every 0.48 ps allow tracking of the structural variations with adequate temporal resolution. For different versions of the 2D histograms with various time shifts, see the Supplementary Material.

At negative delay times, corresponding to the system's evolution at room temperature, the histogram has a single, well-defined maximum located at the origin. The broad and weak "tail" stretched along the diagonal corresponds to atoms belonging to structurally disordered grain boundaries. The off-diagonal, symmetric "patches" (centered around coordinates (2, 6) and (6, 2)) represent atoms in the vicinity of diffusing point defects, which reversibly change the CSP of the involved atoms.

For $E = 0.48$ MJ/kg (corresponding to the experimentally observed melting onset), the initially sharp maximum broadens and shifts along the diagonal up to ~ 2.5, which is attributed to a continuously increasing thermal disorder during lattice heating. In the XRD pattern (see Fig. 8), this process is observed as attenuation of the Bragg diffraction peaks described by the Debye-Waller factor. At the same time, three additional maxima start to develop clearly: a diagonal one around the point with coordinates (7, 7) and two off-diagonal around (2.5, 7) and (7, 2.5). Roughly equal intensities of the three maxima indicate that the flux of atoms from the ordered to the disordered configuration is balanced by the reverse (disordered – ordered) flux. This behavior is consistent with the formation and migration of lattice defects involving the transient emergence of atoms with high CSP. We note that the value of CSP = 7 was found to be typical for atoms surrounding a single vacancy (see Supplementary Material). At the intermediate energy density ($E = 0.70$ MJ/kg), the above trends in the histograms continue at 2.5 ps, but some differences can be noticed at longer times. Firstly, the off-diagonal maxima have different intensities at $\Delta t = 5$ ps (the top left one is more intense than the bottom right one), meaning that the structural rearrangement proceeds overwhelmingly in the order-to-disorder direction. Secondly, the high-CSP (top right) diagonal maximum initially located around (7, 7) continuously shifts close to (10, 10) position. A detailed analysis of the melting transition (see Supplementary Material) indicates that 10 is the most probable value of CSP of an atom belonging to the liquid phase. The high-CSP diagonal maximum becomes dominant at $\Delta t = 10$ ps, suggesting that most atoms are in a permanently disordered environment. The evolution indicates incomplete melting proceeding on the time scale of approximately 10 ps, consistent with the result shown in Fig. 8. When the energy density reaches 1.1 MJ/kg, the structural reconfiguration accelerates considerably. The transient states involving traces of the off-diagonal maxima can be seen only at $\Delta t = 2.5$ ps. At this delay time, the system remains partially crystalline, as evidenced by the presence of Bragg peaks (see Fig.

8). When $\Delta t = 5$ ps and 10 ps the 2D histogram consists of a single, diagonal maximum centered at CSP = 10 equivalent to a completely molten state.

We note that very similar evolution of the CSP has been observed for an MD simulation of a model system of an infinite Pd single crystal undergoing continuous heating. Furthermore, the model system reasonably reproduces the features observed in the experimental XRD patterns, providing an additional support for our findings. For the detailed results obtained for the model system, see the Supplementary Material.

## Discussion

The results of the variation of the crystalline and liquid contributions (integrated diffracted signal of the fcc (111) Bragg peak - $A_C$ and of the liquid peak - $A_L$) presented in Figures 3, 4, and 5 provide insight into the structural evolution of the optically-excited films. The interpretation of $A_C$ data requires a distinction between two dominant physical mechanisms responsible for the reduction of the intensity of the Bragg peaks. The first one is the thermal motion of atoms about their equilibrium positions in the crystalline lattice with the magnitude of the effect expressed in terms of the Debye-Waller factor (DWF). The second contribution is the melting transition involving a simultaneous increase of $A_L$.

Careful analysis of the variation of $A_C$ shown in Fig. 4 suggests that maximum rate of the decrease of the (111) Bragg peak scales linearly with absorbed energy density with no apparent signs of saturation (Fig. 4D). This observation indicates that within the studied energy density range, the crystal-liquid transition remains limited by the heating rate, not the actual melting mechanism. Thus, despite the much stronger coupling between hot electrons and the lattice, the melting mechanism of Pd is similar to that reported for Au where the coupling is weak and heating of the lattice is relatively slow.

While the interpretation of the variation of the Bragg peak intensity is relatively straightforward, the origin of an increasing "broad" contribution, identified by the numerical peak fitting algorithm as $A_L$, is less evident. Apart from the first traces of liquid formed on melting, a weak "broad" signal can easily be confused with other contributions to a diffuse scattering background. In particular, for high lattice temperatures encountered in our experiment, thermal diffuse scattering, compensating for the loss of intensity of the Bragg peaks and caused by inelastic scattering from phonons, make an additional "broad" contribution to the diffraction signal. Besides the thermal effect, lattice defects give rise to the weak, "broad" scattering signal. The displacement field generated by defect sites contributes to diffuse scattering around the Bragg peaks. The effect, known as Huang scattering (64), has been exploited to quantify vacancy concentration in metals from diffuse background close to Bragg reflections (65), (66). It should be noted that for the current data set, the actual contribution of the defects to the diffuse scattering is difficult to quantify. Furthermore, using the peak deconvolution procedure described above, distinguishing between the diffuse scattering (thermal or defect related) and the actual liquid phase scattering (in particular, when the liquid fraction is low) is challenging. Nevertheless, we suggest that an apparent increase of $A_L$, observed for low deposited energy densities $E < 0.8$ MJ/kg, is, at least to some extent, related to increasing concentration of lattice defects. This viewpoint is supported by a detailed analysis of the TTM-MD results, discussed in the following part of this section. It should also be noted that crystal defects and the related diffuse scattering will contribute to a decrease in the Bragg peak intensity. The scale of the effect, however, is not expected to be significant, as compared to the thermal effect.

As discussed above, an unambiguous interpretation of a single diffraction pattern is challenging. Nevertheless, the temporal evolution (i.e., rate of variation) of the different diffraction features provides means to distinguish effects of heating and melting of the crystalline lattice. In our interpretation, the thermal vibrations of atoms are responsible for the initial, ultrafast decrease of $A_C$ dominant at $E < 0.8$ MJ/kg, while melting sets in only when the fluence of the laser pulse exceeds the melting threshold of 0.8 MJ/kg. The rapid decrease of $A_C$ is consistent with the strong coupling between hot electrons and the crystalline lattice of Pd (*44*). Furthermore, the magnitude of the decrease (~25%) is consistent with the expected variation of the diffracted intensity on heating from room temperature up to the melting point under equilibrium conditions (*67*). This value is in good agreement with the melting onset deduced from the initial, rapid drop of $A_C$ in Fig. 5a. It is worth mentioning that in contrast to melting, the magnitude of the Bragg peak area reduction due to the thermal effect strongly depends on the scattering angle (*52*), the two processes could readily be decoupled in a $\Delta t$-scan, given a higher accessible $q$-range.

Besides the integrated diffraction signal, the width $\Delta\Theta_{111}$ of the (111) Bragg-peak carries additional information on the physical state of the crystalline structure. Significantly, no variation of $\Delta\Theta_{111}$ is observed during the initial ~3 ps, which is consistent with the duration of an ultrafast drop of $A_C$. The decrease of the Bragg peak area without change of its width is characteristic of the thermal disorder effect and thus strengthens the interpretation presented in the previous paragraph.

When the instrumental contribution $\Delta\Theta_i$ (determined from the measurement of LaB$_6$ reference) to the peak width is subtracted from the total width, the remaining width (calculated as $\sqrt{\Delta\Theta_{111}^2 - \Delta\Theta_i^2}$) can be attributed either to the finite size of the diffracting domain or distortions of the crystal lattice (microstrain) (*68*). Size and strain components can be separated by the Williamson-Hall (69) or Warren-Averbach (*56*) approaches based on diffraction peak profile analysis. However, for reliable size-strain analysis, both methods require access to several diffraction peaks, in a possibly wide 2θ range. Due to limitations of the scattering angle resulting from the relatively low photon energy of

X-rays available, only two peaks of the fcc Pd phase could be reached in our experiment, with the (200) peak being too weak for a reliable profile analysis. These restrictions exclude the possibility of a reliable separation of the size and strain effect in the collected XRD patterns. Nevertheless, the variation of the (111) peak width with $\Delta t$ and $E$ correlated with $\Delta t$- and $E$-scans of $A_C$ can provide additional insight into the laser-induced melting process.

At first glance, it is tempting to attribute the increase of the peak width to a decrease of the grain size due to melt-fronts propagating from the grain boundaries. However, the existence of a maximum in the $\Delta t$-dependence of $\Delta\Theta_{111}$ suggests that the primary origin of the Bragg peak broadening is a transient increase of a non-uniform strain rather than a grain size variation. Since the strain broadening of the Bragg peak can be regarded as an effect of the superposition of multiple shifted peaks, the variations of $\Delta\Theta_{111}$ and the peak position should be correlated, providing information on an acoustic response of the optically excited film. In our interpretation, the initial increase of the width is mainly due to the non-uniformity of the strain, which builds up for approximately 10 ps, during dynamic lattice expansion, evidenced by the peak shift. The subsequent decrease of $\Delta\Theta_{111}$ arises from the homogenization of the strain and occurs once the expansion is completed and the peak position reaches a constant value. We note that the above explanation of the experimental XRD data is supported by the numerical modeling results discussed in the following section. Also, it should be mentioned that the proposed scenario cannot rule out a contribution of the decreasing crystal size to the observed evolution of the $\Delta\Theta_{111}$. However, the rise of the peak width cannot be considered evidence of heterogeneous melting initiated inside the grain boundaries and propagating into the bulk interior of the grain.

The above discussion highlights the complexity of a quantitative XRD analysis of the short-pulse laser-induced melting transition and exposes intrinsic limitations in the interpretation of the experimental data. It follows that a detailed reconstruction of the solid-to-liquid transformation requires realistic computer simulations at relevant temporal and spatial scales. Our assessment is based on a direct comparison of the experimentally measured diffraction patterns with those calculated from the MD configurations. Besides the fundamental constraints of the simulated diffraction patterns discussed in the previous section, it is worth mentioning that the theoretical patterns do not account for all the details of the experiment (geometrical aspects of the setup, detector characteristics etc.), giving rise to the actual form of the measured XRD patterns. Further limitations involve the MD simulations, which do not reproduce all the specific details of the thin Pd films deposited by magnetron sputtering (details of the polycrystalline microstructure, film-substrate interface etc.). It should also be mentioned that electron-phonon coupling at the disordered grain boundaries can be locally increased due to enhanced electron scattering (*70*). This effect, which might increase the contribution of heterogeneous melting, is difficult to quantify (*21*) and thus not considered in our

TTM-MD simulations. While the above arguments suggest that a perfect quantitative agreement between the experimentally observed and the simulated diffraction patterns can not be expected, our data nevertheless indicate that the agreement is not only qualitative but largely quantitative, which proves the realistic character of the simulations.

According to the current simulation data, optically induced melting of the polycrystalline Pd film is initiated inside the disordered grain boundary regions. Heterogeneous nucleation of the liquid phase in GBs dominates in the low energy density regime, where the film is molten only partially. During the GB melting, the transformation front propagates into the interiors of the grains at a velocity reaching ~400 m/s (13% of the room temperature sound velocity in Pd). This value is in fair agreement with the previously reported data for metals (*21*, *41*, *71*), which suggests that the melting front propagates at 3–12% of the room temperature speed of sound, depending on the local heat dissipation conditions. While the actual value of the melt front propagation velocity is difficult to establish, it is justified to assume that the speed of sound is its upper limit. With the speed of sound in Pd of ~3000 m/s and the grain size of 30 nm (as deduced from the TEM analysis of the as-deposited films), 5 ps is required for the complete melting of a grain via a purely heterogeneous mechanism involving melt front propagating from the GBs into the bulk. According to the current experimental data, at energy densities above 2.0 MJ/kg, the film melts entirely in less than 5 ps from optical excitation. Considering that the above estimation assumes the limiting case of melt front velocity propagating at the speed of sound, it is reasonable to conclude that heterogeneous melting cannot account for the experimental observations and that homogeneous melting must be included in the description of the crystal-solid transformation at high energy densities. This simple estimation is in line with the results of the MD simulations. While the MD snapshots show partial melting starting from the GB regions for low energy densities (Fig. 8), an increasing contribution of homogeneous nucleation of liquid inside the crystalline grains becomes more pronounced in the high-energy-density regime. No clear onset point of homogeneous melting is detected in the MD results, and a smooth transition between the purely heterogeneous and mixed (hetero- and homogeneous) melting is observed.

A detailed analysis of the CSP derived for the MD simulations and the corresponding variations in the diffraction patterns provides additional insight into the relationship between the atomic short- and long-range order (SRO and LRO) during melting. In the current study, the CSP quantifies the SRO while the LRO is probed by diffraction. Since melting is essentially a loss of the long-range crystalline order, and the LRO emerges when the short-range interactions are repeated over infinitely great distances, the variations in SRO and LRO are expected to be correlated. An interrelation between the spatial distribution of the CSP, its temporal evolution, and the diffraction patterns exposes this correlation and reveals a comprehensive picture of the melting transition. The results presented in Fig. 8 and Fig. 9 indicate that the structural pathway for the ultrafast melting of the polycrystalline Pd film involves three distinct processes affecting both the short- and the long-range atomic order.

The first process involves a continuous rise of the CSP—namely occurring via a relatively small variation of the centrosymmetry—from the initial (room temperature) mean value of nearly 0, up to ~2.5 close to the melting onset. This variation of the CSP is related to the uniform heating of the crystalline lattice and the increasing thermal atomic disorder throughout the film. It is manifested in the diffraction pattern by a ~25% decrease of the (111) Bragg peak intensity occurring at a constant peak width.

The second process involves only a fraction of atoms and is a discontinuous, transient increase of the mean CSP up to ~7. This variation is consistent with the development of highly mobile lattice defects, in particular vacancies, occurring simultaneously with the heating process. The defects contribute weakly to the diffraction pattern through local lattice distortions, giving rise to diffuse scattering in the vicinity of the Bragg peaks. Since the position of the (111) Bragg peak of Pd fcc coincides with the position of the broad liquid peak, the diffuse contribution from lattice defects is identified by the peak fitting algorithm as an increase of $A_L$ before the actual melting threshold is reached (see Fig. 5b).

The third process sets in when the hot crystal becomes highly defective and involves a continuous, permanent increase of the CSP up to ~10 – the value characteristic for the liquid phase. The experimental indication of melting is the vanishing of the Bragg reflections and the simultaneous emergence of a broad maximum in the diffraction pattern. According to the current results, the loss of LRO on melting is mediated by local degradation of the SRO due to lattice defects. Those defects can be pre-existing (grain boundaries) or generated during ultrafast lattice heating. The heterogeneous melting can be thus viewed as a slight increase of the local atomic disorder in the grain boundaries followed by propagation of the transformation front into the ordered bulk region. Consequently, homogeneous melting occurs once a certain fraction of atoms in the crystal bulk is involved in a defective local environment, i.e., when a critical concentration of defects is reached. We note that a similar criterion for melting has been proposed in the early works of Górecki (*11*). According to Górecki's model, melting starts when the vacancy concentration in solid metal reaches a threshold value of 0.37%, and melting involves a discontinuous jump in vacancy concentration up to 10%. A defect-induced melting mechanism has also been proposed by Fecht (*16*). He suggested that above a certain degree of frustration, expressed in terms of a critical defect concentration of 7.7%, the crystal becomes unstable against liquid-like heterophase fluctuations, and the system encounters an intrinsic instability (entropy catastrophe). More recently, the role of crystal defects in the melting process was experimentally demonstrated by

Mo et al. (*32*). Their study shows that radiation-driven defects in tungsten, in particular their clusters, lower the melting temperature during an ultrafast solid-liquid phase transition.

A detailed analysis of an ideal, infinite single crystal MD system which melts via a purely homogeneous mechanism shows that in the superheated regime, near the melting point, the populations of atoms forming a hot crystal (CSP of ~2.5) and atoms involved in a defective environment (CSP of ~7) are comparable (deduced from the height of peaks at sub-melting point temperature in Fig. S9). Since an isolated point defect affects all its 12 nearest neighbors in the crystalline fcc lattice, the actual concentration of defects is at least an order of magnitude lower than that of the atoms with a defective environment. According to this rough estimation, the concentration of defects in crystalline Pd at its homogeneous melting threshold is of the order of several percent. This value lies in the range between the value predicted by Górecki for the melting temperature and that proposed by Fecht as an ultimate stability limit of a crystal.

An important aspect of the melting process is the role of the strain waves generated by the fast heating of the film. Given the planar geometry, out-of-plane expansion is more likely than in-plane expansion and thus one can expect 1D expansion in the film's normal/X-ray beam direction. Taking the experimentally observed (111) peak shift of ~0.7 deg and assuming a 1D expansion (*57*), the resulting strain exceeds 20%, which is far beyond any realistic value for metallic solids. On the other hand, if 3D expansion is assumed, the resulting strain reaches approx. 2%. Given that the volume increase of Pd due to 3D thermal expansion on heating from the room temperature to $T_m$ is approximately 6% (*72*), the corresponding lattice spacing would increase by one third of this value – consistent with the experimentally observed peak shift of 2%. It must be noted, however, that our current experimental data do not allow a conclusive distinction between the 1D and 3D expansion scenario since our analysis is restricted to a single Bragg peak. With a higher $q$-range (and thus higher-order Bragg peaks) accessible, the strain geometry could be unambiguously resolved using the approach described in ref. (*57*). The analysis of the MD configurations to resolve the strain is also non-conclusive due to periodic boundary conditions in the in-plane directions, making the strain intrinsically 1D.

While this study focuses on the melting dynamics and not the acoustic response, we can conclude that the strain evolution in the optically excited films might be non-trivial.. Moreover, the current MD results support the interpretation of the peak broadening being caused by a non-homogenous strain distribution inside the film resulting from strain waves launched at the film boundaries and propagating into the bulk of the film. According to our MD simulations, large tensile stresses are generated when these strain waves overlap. For sufficiently high energy densities this can lead to void formation, fracture, and eventually to ablation of the film. (*73*, *74*). One may speculate that tensile stresses in combination with the increasing lattice temperature contribute to the development of excess lattice defects in the crystalline film, accelerating the melting transition. In fact, tensile stress and resulting lattice distortions and stress gradients were found to lower the melting point in MD-simulated Ni metal (*75*). A similar scenario cannot be entirely excluded in the current case of Pd. We note, however, that careful analysis of the MD snapshots does not provide evidence for a correlation between local stress and melting. Moreover, our results indicate that at high energy densities, homogeneous melting of the film occurs before the strain waves have propagated fully into the film. Therefore, we conclude that while tensile stresses may affect the crossover between heterogeneous and homogeneous melting, the local pressure variations in the irradiated film seem to play only a secondary role in the observed melting transition.

Finally, we point out that we did not observe the formation of intermediate diffraction peaks as reported for Au by Assefa *et al*. (*39*), which the authors attributed to material trapped between the solid and melted state. However, the films investigated in the present work were considerably thinner (30 nm) than those studied in ref. (*39*), where the films were 50, 100, and 300 nm thick. According to ref (*39*), splitting of the peak into two distinct components was most pronounced in the 300 nm film and impossible to resolve in the 50 nm one. Therefore, the lack of any intermediate diffraction peaks in the currently studied samples is not unexpected.

## Summary

To summarize, we presented a comprehensive experimental and theoretical analysis of optically induced ultrafast melting of polycrystalline thin films of Pd. The discussed physical picture is based on the results of time-resolved X-ray diffraction experiments on the photoexcited films and large-scale two-temperature model molecular-dynamics simulations performed on the relevant spatial and temporal scale. Our results demonstrate that with increasing density of deposited energy, the melting mechanism progressively varies from relatively slow, heterogeneous melting, initiated inside the material at the structurally disordered grain boundaries, to homogeneous melting, which proceeds catastrophically in the whole crystal volume on a few picosecond time scale. The experimentally evidenced ultrafast melting in Pd is faster by at least a factor of five than in Au (*41*), where the electron-phonon coupling is weak and thermal equilibration is slow. Within the studied energy range, the experimental data do not show a sign of saturation of the melting rate. The rate of the crystal-liquid transition in Pd remains controlled by the lattice heating process despite the much faster energy transfer from hot electrons compared to Au. We thus conclude that the laser-induced melting in Pd is a thermal process, like in other metals investigated so far (*32*, *41*), with its rate limited by the energy exchange and thermal equilibration between the excited electrons and the lattice.

Moreover, thanks to the possibility of the direct comparison of the experimental data with MD simulations, our results emphasize the role of crystal defects in the melting process and correlate the variations in the atomic short- and long-range order. We demonstrate that as the optically excited film's temperature rises, the defects' concentration gradually increases. The melting transition – experimentally characterized by the loss of long-range order, as evidenced by vanishing Bragg diffraction peaks – occurs when a hot, highly defective crystal transforms into a homogeneous liquid due to an abrupt decrease of the short-range order. In this picture, the grain boundary regions melt prior to the bulk due to their pre-existing structural disorder, which facilitates their promotion to the highly disordered liquid state.

Experimental XRD data in a higher $q$-range are now needed to decouple the thermal and melting effects contributing to the decrease of the Bragg peaks' intensity and fully resolve the acoustic response of the laser-excited films.

## Acknowledgments

The authors thank Professor Jerry Hastings for valuable discussion and advice on design of the experiment at the Scientific Instrument FXE at the European XFEL. We acknowledge European XFEL in Schenefeld, Germany, for provision of X-ray free-electron laser beamtime at the Scientific Instrument FXE (Femtosecond X-Ray Experiments) and would like to thank the staff for their assistance.

**Funding**:

This work was supported by the Materials Technologies project granted by:

Warsaw University of Technology under the program Excellence Initiative: Research University (ID-UB), the National Science Centre, Poland, grant agreement No 2017/27/B/ST3/02860,

Deutsche Forschungsgemeinschaft (DFG, German Research Foundation) through the Collaborative Research Centre 1242 (project number 278162697, project C01 *Structural Dynamics in Impulsively Excited Nanostructures*).

The access to the European XFEL was supported by a grant of the Polish Ministry of Education and Science - decision no. 2022/WK/13.

We also acknowledge the usage of the computer cluster DWARF at Warsaw University of Technology supported by the Polish National Science Center (NCN) under Contracts No. UMO-2017/26/E/ST3/00428.

I.M. gratefully acknowledges financial support from Dutch Research Council (NWO) (Project 'PROMT', Grant Rubicon Science 2021-1 S, file number 019.211EN.026), and the Industrial Partnership Program 'X-tools', project number 741.018.301, funded by the Netherlands Organization for Scientific Research, ASML, Carl Zeiss SMT, and Malvern Panalytical. This work was (partially) carried out at ARCNL, a public–private partnership of the UvA, VU, NWO, and ASML.

**Author contributions:**

Conceptualization: JA, KST, PZ, RS

Methodology: JA, AO, KST, PZ, IM, PDz, VVZ, RS

Investigation: JA, AO, KST, PZ, IM, PDz, CB, MC, PDł, ARF, KF, WG, KG, ALG, IJ, RWEK, RK, DKh, DKl, KMK, KK, KPM, RM, NTP, MS, PS, HY, VVZ, RS

Writing - Original Draft: JA, KST, PZ, IM, RS

Writing – Review & Editing: JA, AO, KST, PZ, IM, HNC, ARF, KG, ALG, RWEK, RK, DKl, MS, PS, RS

Supervision: JA, KST, PZ, RS

Visualization: JA, AO, PDz, IM, WZ, RS

Funding acquisition: JA, KST, IM, PDz, RWEK, RS

**Competing interests**: The authors declare that they have no competing interests.

**Data and materials availability**: All data needed to evaluate the conclusions in the paper are present in the paper and/or the Supplementary Materials.

# Supplementary Materials for

**Structural pathways for ultrafast melting of optically excited thin polycrystalline Palladium films**

Jerzy Antonowicz*, Ryszard Sobierajski* et al.

*Corresponding author. Email: jerzy.antonowicz@pw.edu.pl, ryszard.sobierajski@ifpan.edu.pl

**Supplementary Text**

Samples

Figure S1A shows a schematic cross-section of a sample consisting of X-ray transparent $Si_3N_4$ windows etched on Si wafer frame and a thin Pd layer capped with $SiO_2$ with schematically indicated pumping (optical) and probing (X-ray) pulses. As indicated in the figure, the samples were irradiated from the cap side. An optical microscopy image of the selected area of the sample, seen from the Si frame side (Fig. S1B), shows nine windows after irradiation with a single X-ray pulse. The bright spots in the center of the windows are holes generated by the X-ray pulse, which locally destroys the film.

The cross-section TEM micrographs of the as-deposited films are shown in Figure S2.

Experimental setup of the pump-probe experiment

Figure S3A shows a schematic layout of the pump-probe experiment at the FXE instrument of the EuXFEL involving the acquisition of the X-ray diffraction patterns in transmission geometry. The vacuum chamber is pictured in Fig. S3C, and Fig S3C shows the inside of the chamber with the sample frames mounted on a sample holder. The exit flange covered with Kapton window faces the LPD detector.

XRD data analysis

*Background subtraction and normalization*

In order to extract reliable structural information on the metallic film, a "dark" background signal ("dark current" of the detector) as well as the scattering signal from the SiN substrate and the $SiO_2$ cap, has to be subtracted from the measured data. Furthermore, due to the stochastic nature of the self-amplified spontaneous emission (SASE) X-ray beam at the EuXFEL and the resulting fluctuations of the X-ray pulse energy, the data require proper normalization. Before normalization of the azimuthally integrated patterns, an averaged background signal was subtracted from the data. To eliminate the contribution from the SiN and substrate/cap layers, specially prepared "bare" samples were prepared and measured in course of the experiment. Those samples consisted of Si frames with 300 nm $Si_3N_4$ membrane capped with 300 nm of $SiO_2$ – identical to the proper samples, with the only difference being the lack of the metallic layer.

Both averaged background signals were removed from the data gathered from the actual Pd sample films. First the "dark" background was subtracted before any normalization, since its intensity was identical for each measurement. The "bare" sample signal had to be scaled to each pattern individually before subtraction. This was achieved by multiplication of the Pd sample pattern by a factor yielding a maximum overlap of the two patterns in the 10-15 deg 2θ range. This range was selected for the following reasons: (i) neither crystalline nor liquid Pd produces a diffraction signal between 10 and 15 deg, and thus any signal originates from the $Si_3N_4$ and $SiO_2$ layers, (ii) the range is free from incidental spikes in the "dark" background (see Fig. S4A). Finally, the resulting patterns were normalized to maintain the same total intensity of each pattern in the full 10-50 deg range. The result of the background subtraction and signal normalization is presented in Figure S5.

For further analysis, highly noisy or flawed patterns (taken at defective sample windows or with exceptionally low energy X-ray pulses) were rejected and excluded from further analysis.

*Peak fitting*

Peak fitting of the azimuthally-integrated XRD patterns was performed in MATLAB in the 2θ range of 20 – 45 deg, covering the angular region of the (111) and (200) Bragg reflections of the Pd fcc crystalline phase and the broad peak originating from the liquid phase. Prior to fitting, a linear background was subtracted from the data. Each fit involved four peak functions. Three Pseudo-Voight (50/50 Gaussian/Lorentzian contribution) functions represented the (111) and (200) peaks of Pd fcc and a weak peak being a (111) reflection from the non-excited ("cold") part of the film (for example, see Fig. S4). A Gaussian peak represented the liquid contribution to the diffraction pattern.

The values of peak parameters taken for further analysis were obtained by averaging several measurements taken at equal values of the pump-probe delay time and absorbed energy density.

*Refinement of the pump-probe delay time*

During the experiment the pump-probe delay time had to be calibrated by changing the optical aperture parameters, so that at delay time set to 0 both X-ray and optical pulses reach the sample simultaneously. However, over the time of the experiment this calibration exhibited gradual degradation over time, resulting in the drift of $t_0$ – a set delay time value at which both pulses actually hit the sample at the same time. Data shows $t_0$ changed between around -4 and +2 ps set delay time over the course of the experiment. Because of that the delay time had to be refined for each delay run individually.

The following approximations were assumed to achieve this refinement: a) pump pulse had a Gaussian profile with 900 fs FWHM, b) after interaction with the sample the temperature of the film changes according to the two-temperature model (TTM), where it follows a shape of an exponentially modified gaussian distribution - a convolution of the normal and exponential distributions. In this model the gaussian contribution comes from the pump-laser so it has the same 900 fs FWHM. and lastly c) the $t_o$ is the time at which the intensity of the gaussian component first reaches half of its maximum value.

Both position and height of the (111) Bragg peak change with temperature, and because of that all parameters of the fitted (111) peak: position, height and area were considered for this procedure. However, after testing with both experimental and MD simulated data the (111) peak height was ultimately selected. Fig. S6 shows an example of a $t_0$ refinement using the expnormal distribution fitted to change of (111) peak height. After finding a refined $t_o$ for each delay run all delay time in respective runs were shifted accordingly. Figure S7 shows an example of fitted data before and after the $t_o$ refinement.

Molecular dynamics simulations
*EAM potential development*

Atomistic simulation of melting with the molecular dynamics (MD) method requires an appropriate interatomic potential reproducing the experimental melting temperature $T_m$ with a good accuracy. The available embedded atom model (EAM) potentials (*76, 77*) for Pd provides $T_m$ = 1583 K and $T_m$ = 1552 K, respectively, which is notably below the experimental $T_m$ = 1828 K. For realistic MD simulation, we developed a new EAM potential using the stress-matching method (*78, 79*).

Such development proceeds through the following well-defined steps. The first step involves generating a database of the experimental data extrapolated to zero temperature (e.g., lattice size, elastic constants, cohesion, and surface energies), and the first-principle cold stress tensor curves obtained by the density functional theory (DFT). The second step involves fitting the EAM functional parameters to the database, which produces several good EAM parameter sets (candidates). For the last step, which requires MD simulations, the best potential is selected from the candidates depending on how well they reproduce important experimental properties not included in the fitting, such as the melting temperature and the surface tension.

As in the original work (*78*) the fitting database contains the stress tensor components $\sigma_{\alpha\beta}(V) = -p_{\alpha\beta}(V)$ in the palladium *fcc* lattice under continuous hydrostatic compression up to $p = 800$ GPa and stretching by a factor 1.6, and uniaxial deformation along the [100]-axis calculated by DFT VASP (*80*) package using the PAW method (*81*). The grid of $25 \times 25 \times 25$ of Monkhorst-Pack points together with the cut-off energy of 680 eV were used to reproduce electronic structure of Pd valence bands consisting of 18 electrons with the help of a GW pseudopotential. For verification of the obtained single-electron energies, the all-electron method FP-LAPW implemented in the computational code Elk (*82*) was used. In both all-electron and pseudopotential calculations, the exchange-correlation functional was applied in the GGA approximation (*83*).

The fitting procedure implies the constraints of the monotonic behavior of $p(V)$ including the requirement that the sound velocity increases with the compression.

The database also includes the known cohesive energy of 3.91 eV, the bulk modulus, and elastic constants extrapolated to $T = 0$ K (*77, 84*).

The lattice parameter of $a = 0.38733$ nm at $T = 0$ K was set to provide the experimental density 12.023 g/cc of Pd at room temperature. Apart from the above, the unrelaxed vacancy formation energy of 1.59 eV, the stacking fault energy of 0.151 J/m², and the surface energy of solid $\gamma_s(111) = 1.61$ J/m² are added to extend the database, see those energies also in (*77*). It is worth mentioning that, since the surface energy $\gamma_s(111)$ of solid is not well-known, it was adjusted several times in fitting the intermediate potential candidates to obtain better agreement with the experimental surface tension of liquid palladium at the melting point.

The phase coexistence method is used to calculate the melting temperature at nearly zero pressure. It is found that the new EAM potential provides $T_m = 1835$ K (using the phase coexistence method (*85*)) and liquid density $\rho = 10.72$ g/cc, which are close to the experimental data $T_m = 1828.05$ K and $\rho = 10.66$ g/cc. MD simulation of an equilibrium liquid--vapor system with a flat interface is used for the calculation of the surface tension. At the melting temperature, our new EAM potential yields $\sigma = 1.47$ J/m², which is close to the experimental $\sigma = 1.48$ J/m² (*86*).

The newly developed potential is available in both a rational function set and in the tabulated form for the LAMMPS simulation package.

All MD simulations are performed with our in-house parallel code MD-VD$^3$ utilizing the Voronoi dynamic domain decomposition (*78, 87*).

*Comparison with the experiment*
In order to correlate the experimental and the MD-simulated temporal dependencies of the normalized area o the (111) Bragg peak, $A_C$, we introduced a numerical procedure to find the best match between the experiment and the theory. Given a data set consisting of seven experimental and eight theoretical (resulting from the TTM-MD simulations) $\Delta t$ dependencies of $A_C$ (Figure S5) for each experimental $E$ ($E_{\text{exp}}$), we systematically evaluated the mean squared error (MSE) for each of the eight experiment-theory pairs and plotted the MSE versus the theoretical energy density, $E_{\text{sim}}$. An example of this procedure is shown in Figure S6 for $E_{\text{exp}} = 1.3$ MJ/kg. The best match between the experiment and the theory occurs at the location of the minimum of the function MSE ($E_{\text{sim}}$). To refine the location of the minimum, we approximated the MSE ($E_{\text{sim}}$) with a fifth-order polynomial and evaluated its double roots. As shown in Figure S7, the above procedure yields the correlation between the experimental and the theoretical densities of the absorbed energy.

*Model Pd system*

Figure S8 presents the evolution of the total system energy during heating ($10^{13}$ K/s) of a model system consisting of 32 000 Pd atoms arranged in a single crystal configuration with periodic boundary conditions in three dimensions. The system melts homogenously at 2420 K. The corresponding temperature-dependent CSP histograms are shown in Figure S9. Figures S10 and S11 present the evolution of the pair distribution functions and simulated X-ray diffraction patterns.

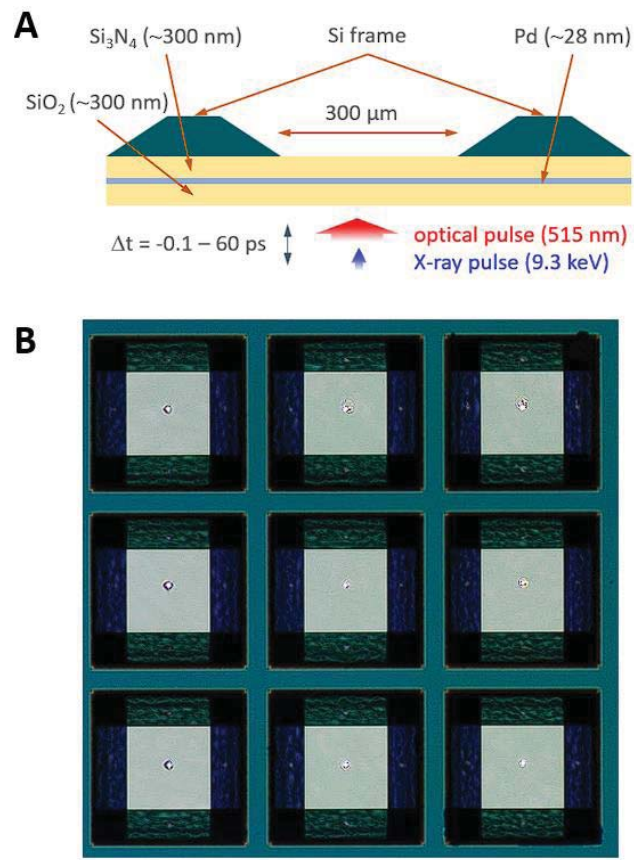

**Fig. S1. Sample layout.** Schematic cross-section of the thin film sample and pumping/probing pulses (**A**) and an optical microscope image of nine sample windows (as seen from the Si frame side) after irradiated by X-rays (**B**). Bright spots in the center of the windows are holes due to local destruction by X-ray pulses.

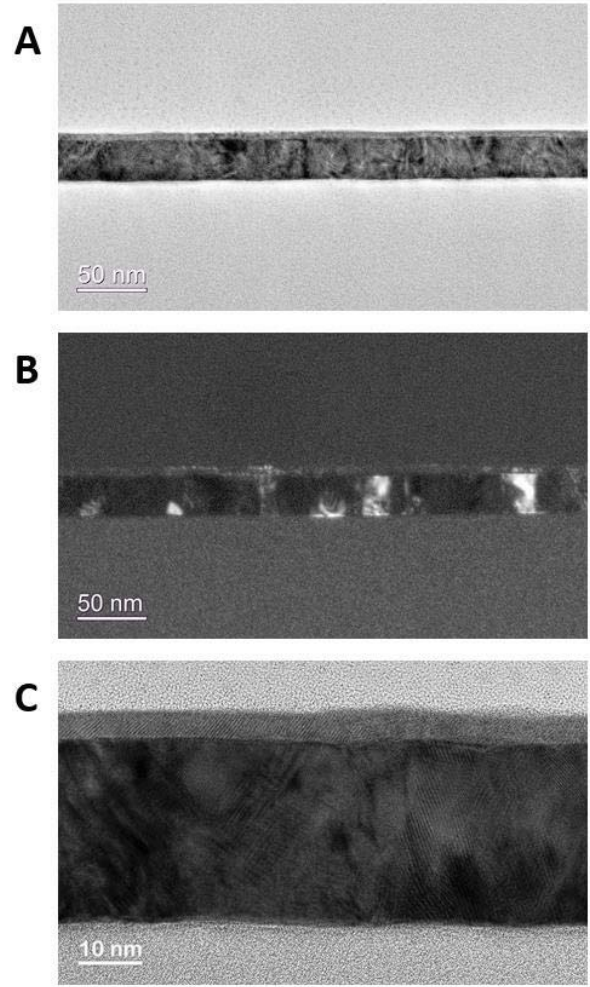

**Fig. S2. The cross-section TEM images of the as-deposited films.** The bright-field image (**A**) of SiN substrate, Pd film, and SiN cap layer are visible. The dark-filed image (**B**) obtained for Pd (111) diffracted beam shows the size of Pd nanocrystals, and the high-resolution image (**C**) picturing the details of the Pd grains structure and an interfacial layer between Pd and the SiN cap layer**.**

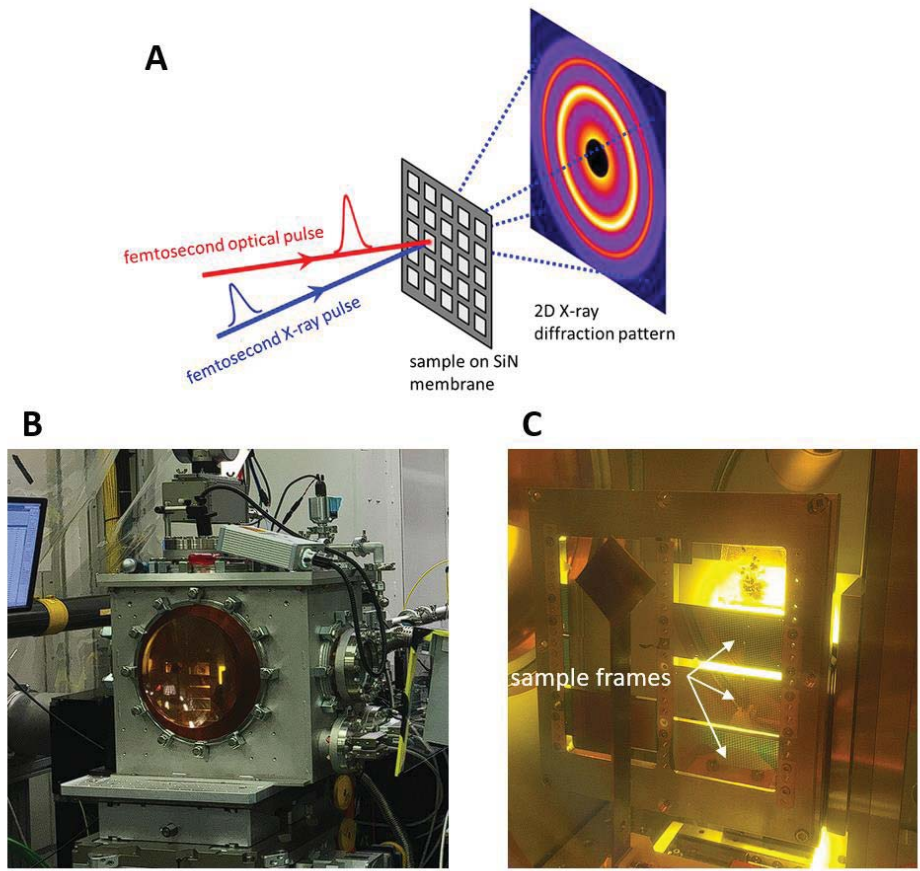

**Fig. S3. Experimental geometry and setup.** Schematic layout of the pump-probe experiment (**A**), and pictures showing the external (**B**) and the internal (**C**) view of the sample chamber.

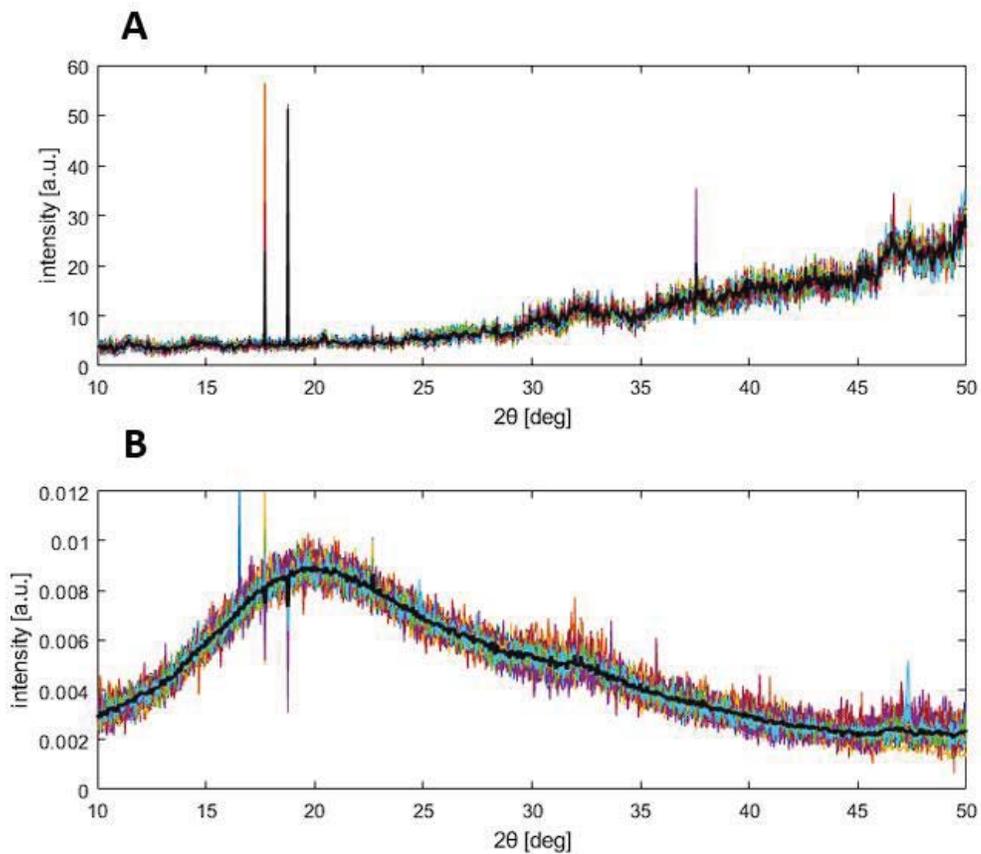

**Fig. S4. "Dark and background signals"**. "Dark" background signal (**A**) and diffraction pattern of "bare" (substrate/cap only) sample (**B**) after "dark" background subtraction and normalization. Both graphs involve multiple X-ray exposures (different line colors), and thick black lines represent the average curve.

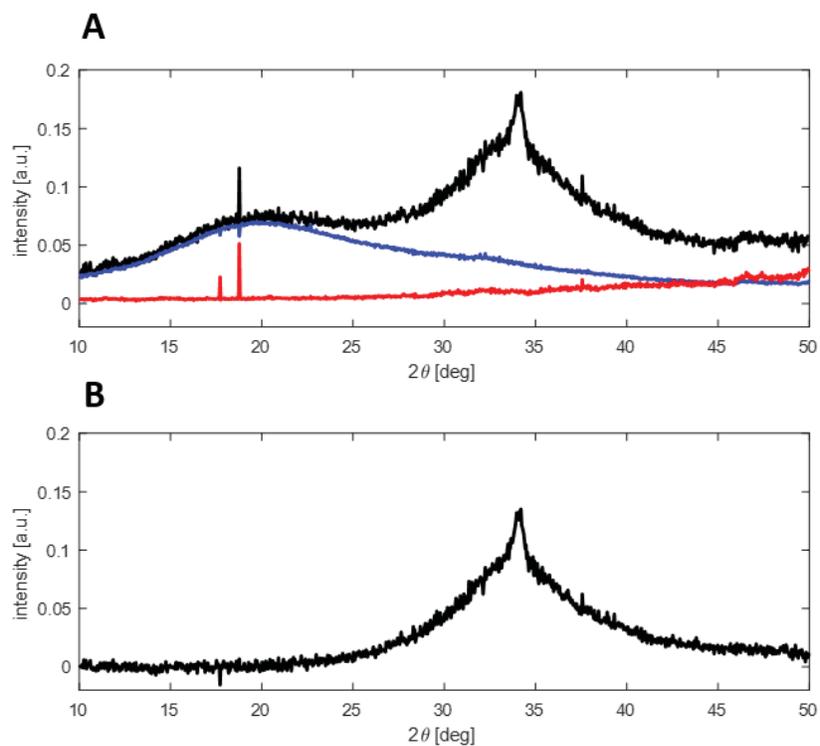

**Fig. S5. "Dark" signal and background subtraction**. A diffraction pattern of a Pd film (black), a "bare" sample (blue), and a "dark" background (red) (**A**). A diffraction pattern of the Pd film after background subtraction and normalization is shown in (**B**).

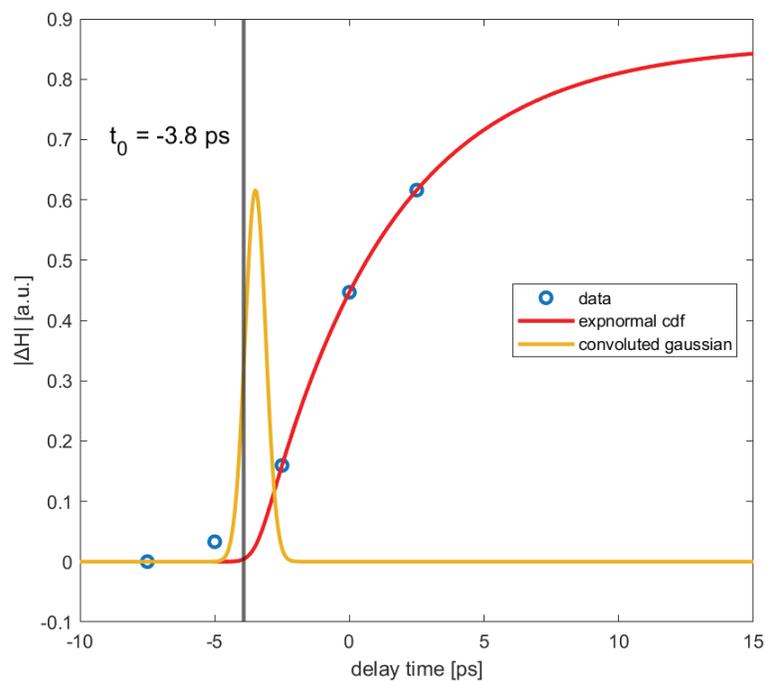

**Fig. S6. Delay time refinement.** The $t_o$ refinement procedure performed using the absolute value of the fitted (111) peak height change in time (blue points). The fitted expnormal function is shown in red, the convoluted Gaussian component at 0.9 fs FWHM in yellow, and the time $t_0$ after refinement is marked with a black line.

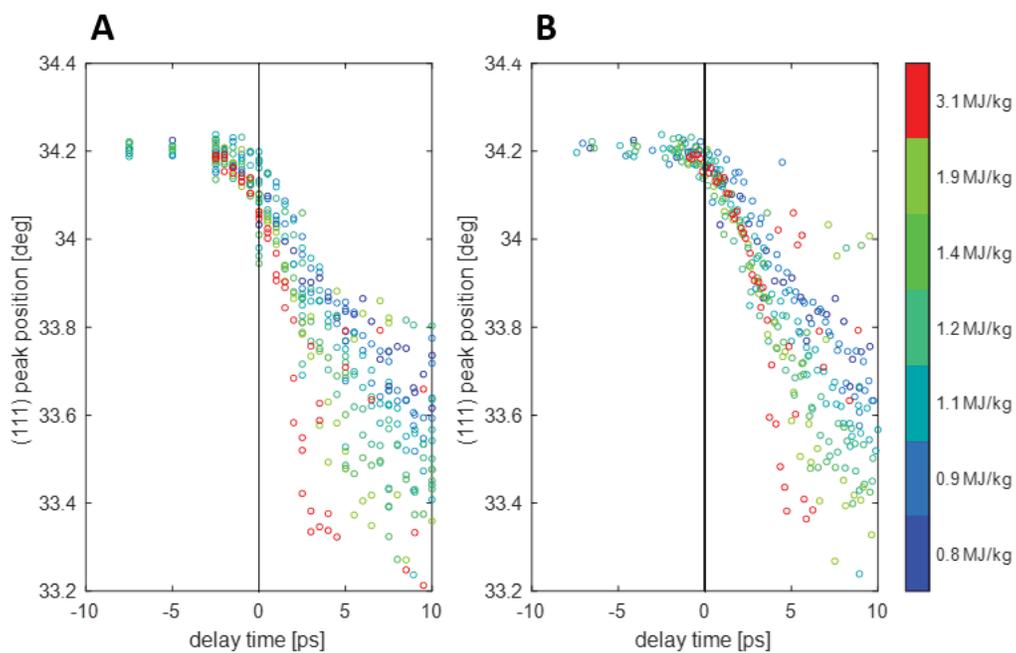

**Fig. S7. Effect of the delay time refinement on the peak fitting results**. The fitted (111) peak position for all delay runs at small delay times before (**A**) and after (**B**) the refinement of $t_0$.

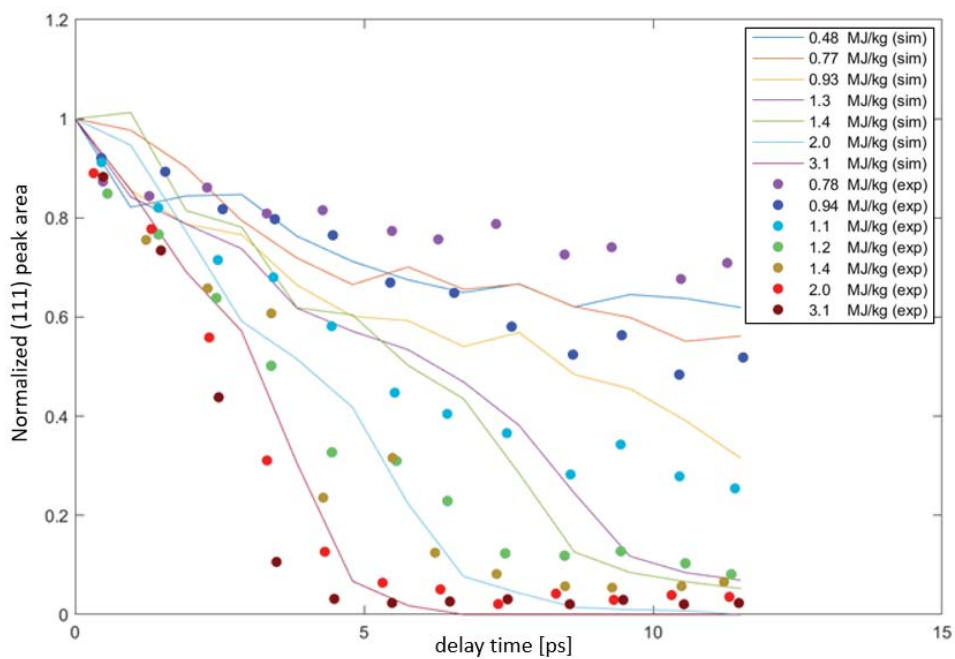

**Fig. S8. Matching the experimental and simulated peak areas**. Experimental (symbols) and theoretical (lines) temporal dependencies of the normalized (111) Bragg peak area.

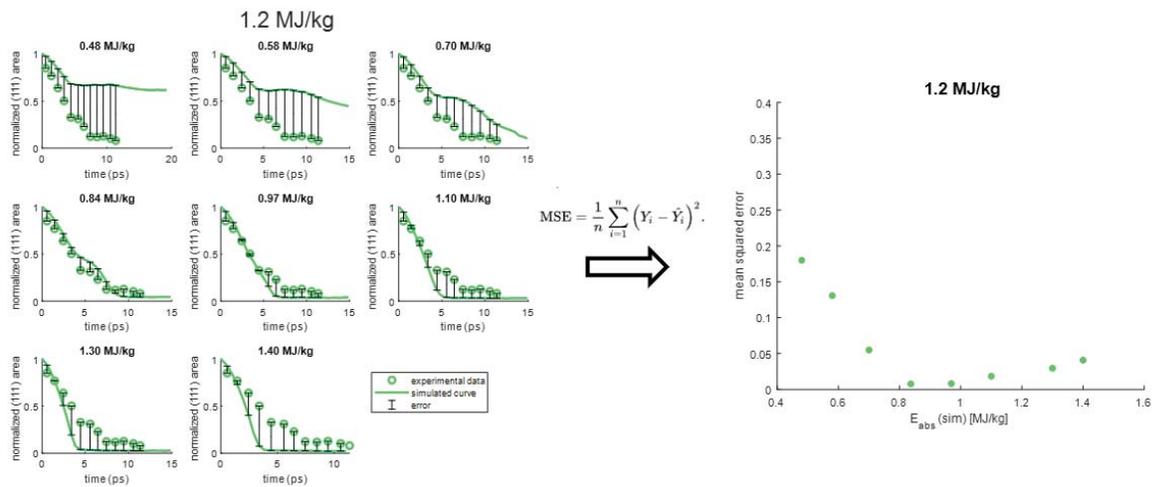

**Fig. S9.** An exemplary evaluation of the MSE for experimental energy density.

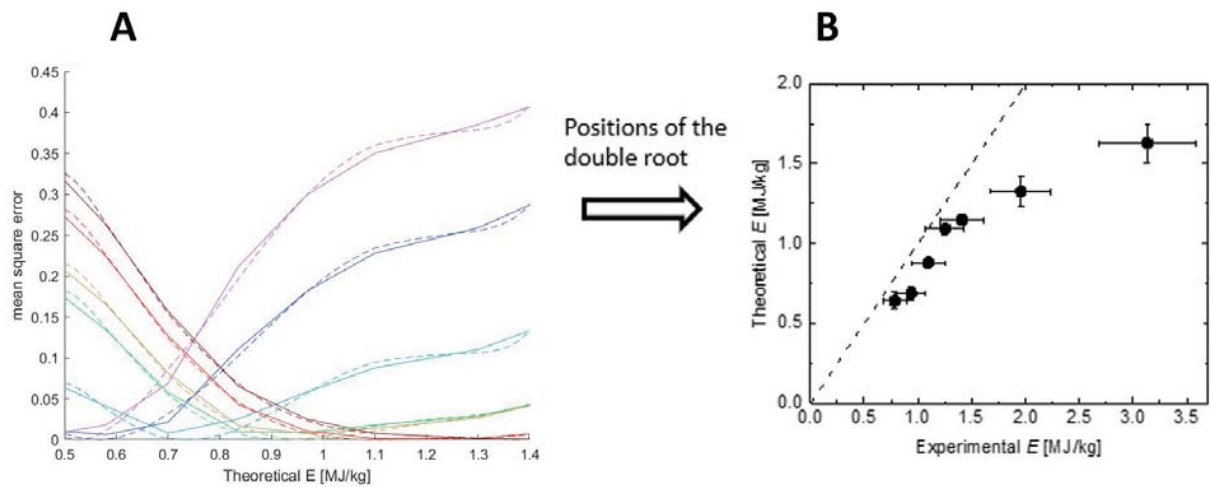

**Fig. S10. Matching of the theoretical and the experimental energy densities**. MSE curves (solid lines) fitted with a 5th-order polynomial (dashed lines) are shown in (**A**). The correlation between the theoretical and the experimental energy densities is shown in **B**. The dashed line in **B** represents a one-to-one correlation.

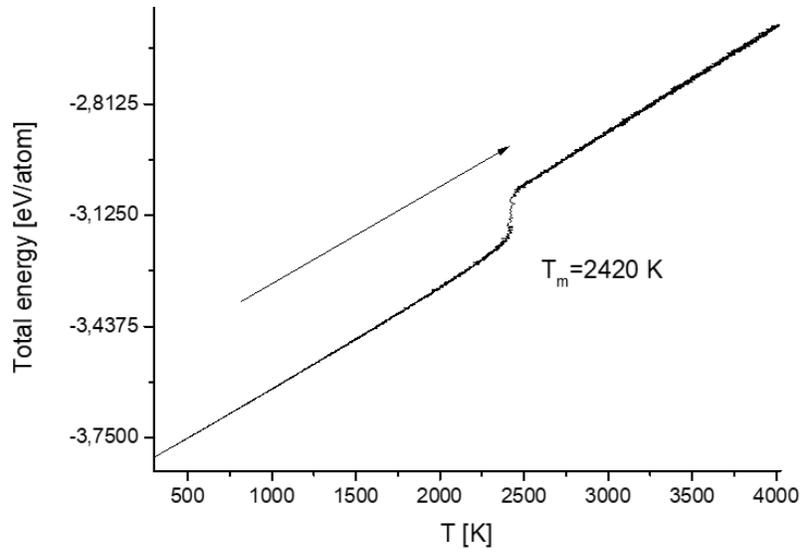

**Fig. S11. Melting of the model Pd system**. Evolution of the total system energy during heating ($10^{13}$ K/s) of a model system consisting of 32 000 Pd atoms arranged in a single crystal configuration with periodic boundary conditions in three dimensions.

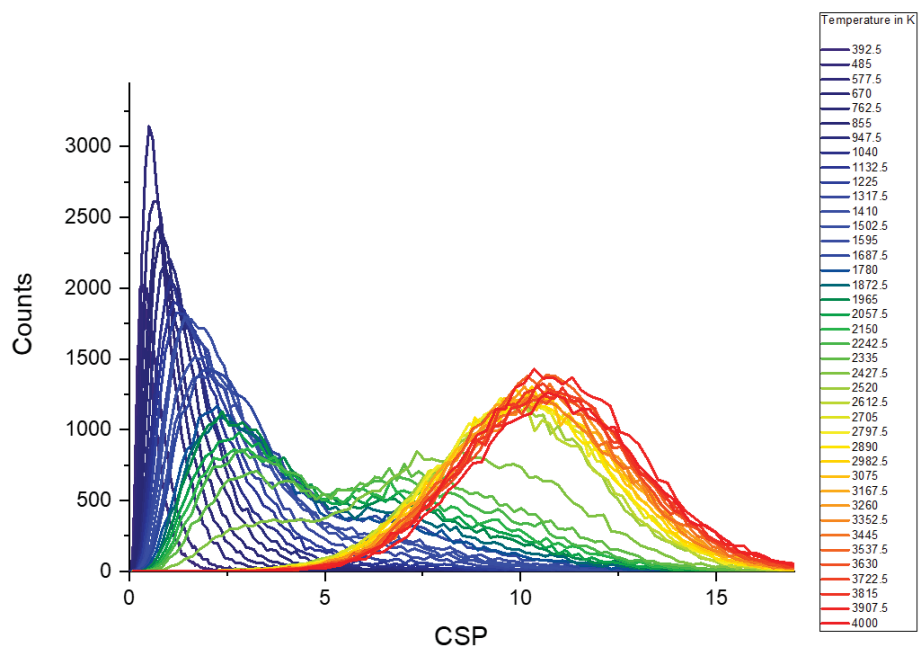

**Fig. S12. Temperature-dependant histograms of the CSP of the model Pd system**.

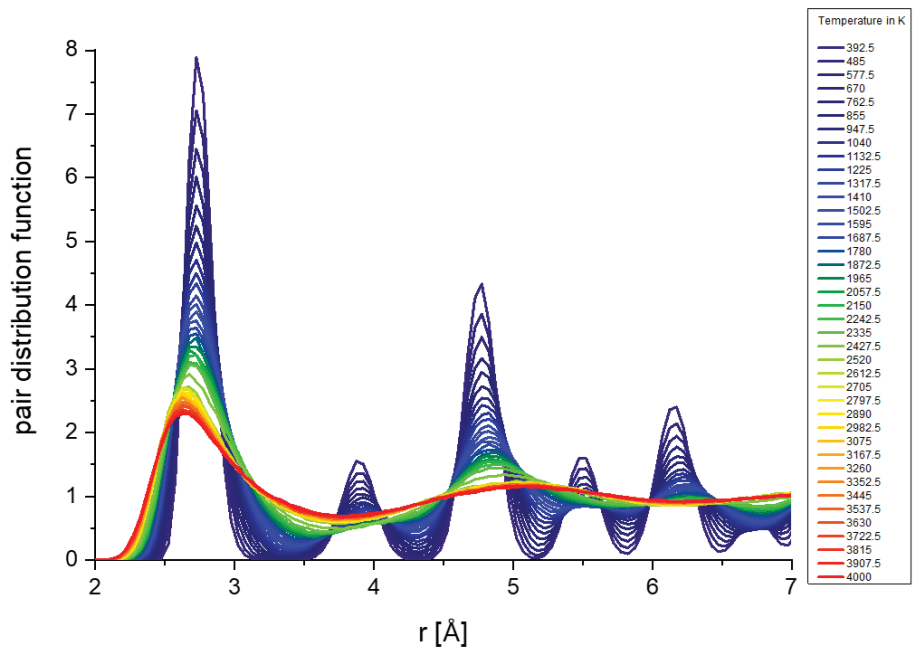

**Fig. S13. Temperature-dependant pair distribution functions of the model Pd system**.

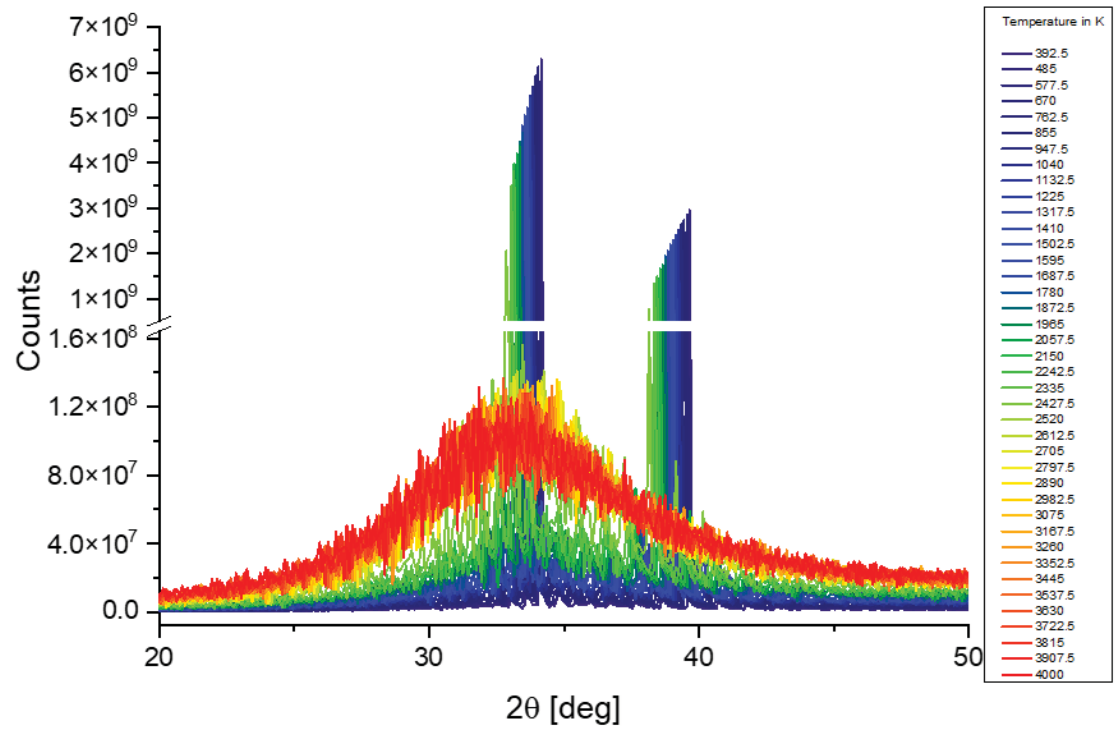

**Fig. S14. Temperature-dependent simulated XRD patterns of the model Pd system**.